\documentclass[toc,cits]{PoS}
\pdfoutput=1

\usepackage{amstext,amsmath,amssymb,amsfonts,bbm}
\usepackage{braket} 
\usepackage{slashed} 
\usepackage{amsthm}  
\usepackage{physics}

\newcommand{\be}{\begin{equation}}
\newcommand{\ee}{\end{equation}} 

\newcommand{\f}{\frac}
\newcommand{\p}{\partial}

\let\a=\alpha \let\b=\beta  \let\g=\gamma  \let\d=\delta
\let\z=\zeta        \let\l=\lambda
\let\m=\mu              \let\r=\rho 
    \let\vph=\varphi  
    
\let\G=\Gamma    \let\L=\Lambda \let\X=F
    \let\Si=\Sigma     
  \let\eps=\epsilon

\newcommand{\phib}{\bar{\phi}}
\newcommand{\psib}{\bar{\psi}}

\newcommand{\cB}{\mathcal{B}}

\newcommand{\cF}{\mathcal{F}}
\newcommand{\cG}{\mathcal{G}}

\newcommand{\cI}{\mathcal{I}}
\newcommand{\cJ}{\mathcal{J}}

\newcommand{\gt}{\tilde{g}}

\DeclareMathOperator{\im}{\mathrm{i}}

\newcommand{\mba}{\mathbf{a}}
\newcommand{\mbb}{\mathbf{b}}
\newcommand{\mbc}{\mathbf{c}}
\newcommand{\mbd}{\mathbf{d}}

\newcommand{\mbm}{\mathbf{m}}
\newcommand{\mbn}{\mathbf{n}}

\theoremstyle{remark}

\title{Melonic CFTs}

\ShortTitle{Melonic CFTs}

\author{\speaker{Dario Benedetti}%\thanks{}
   \\
        CPHT, CNRS, Ecole Polytechnique, Institut Polytechnique de Paris, Route de Saclay, \\ 91128 Palaiseau, 
 France\\
        E-mail: \email{dario.benedetti@polytechnique.edu}}

\abstract{
The melonic limit is a relatively new type of large-$N$ limit, differing from the much older and well-known large-$N$ limits of vector and matrix field theories, which are dominated by cactus and planar Feynman diagrams, respectively. The melonic limit typically appears in tensor field theories, characterized by an invariance group in which the fields transform as the product of $r\geq 3$ fundamental representations of $r$ different simple Lie groups.
As the name suggests, in such a limit  the perturbative expansion of free energy and correlators are dominated by melonic diagrams. The latter form a manageable subset of the planar diagrams, but with a richer structure than cactus diagrams, and therefore they open the possibility of studying in a controlled manner new types of fixed points of the renormalization group. We call \emph{melonic conformal field theories (CFTs)} those fixed-point theories that are found in the melonic limit.
We concisely review the construction and analysis of tensor field theories in $d\geq 2$ (Euclidean) spacetime dimensions, with special emphasis on the general theoretical framework, and on specific results for the fixed points of some models.
}

\FullConference{Corfu Summer Institute 2019 "School and Workshops on Elementary Particle Physics and Gravity" (CORFU2019)\\
		31 August - 25 September 2019\\
		Corfù, Greece}

\begin{document}

%%%%%%%%%%%%%%%%%%%%%%%%
\section{Introduction}
%%%%%%%%%%%%%%%%%%%%%%%%

%The discovery that tensor models admit a novel large-$N$ limit

Large-$N$ limits can provide a helpful approximation scheme for field theories, for example allowing us to gain control over non-trivial fixed points of the renormalization group \cite{Wilson:1972cf} or over non-perturbative phenomena such as spontaneous symmetry breaking and dynamical mass generation \cite{Coleman:1974jh,Gross:1974jv}.  The main feature of the large-$N$ limit is a truncation of the set of Feynman diagrams of a given theory: the large-$N$ limit is dominated by cactus diagrams in typical vector theories (e.g.\ \cite{Coleman:1985rnk}), and by planar diagrams in typical matrix theories  \cite{'tHooft:1973jz,Brezin:1977sv}. Both cases have a long history, with applications ranging from statistical mechanics to QCD and quantum gravity (see for example the collection of articles in \cite{Brezin:1994eb}). More recently, there has been a renewed interest in the large-$N$ limit of theories based on fields with $r\geq 3$ indices, each index ranging from 1 to $N$ and transforming in the fundamental representation of a Lie group such as $O(N)$, $U(N)$, or $Sp(N)$. Zero-dimensional versions of such \emph{tensor field theories},\footnote{The term tensor field theory has been used sometimes for tensor models with a Kontsevich-type free covariance \cite{BenGeloun:2011rc,BenGeloun:2012pu,BenGeloun:2012yk}, otherwise known also as tensorial group field theories \cite{Carrozza:2012uv,Geloun:2014kpa,Lahoche:2015ola,Benedetti:2014qsa,Lahoche:2018hou}. In such models there is no spacetime at the outset, the renormalization group slicing is performed on the tensor indices, which can be thought of as momentum components of some abstract space, in a similar spirit to \cite{Brezin:1992yc}. In contrast, the tensor field theories we will discuss here are ordinary field theories on $d$-dimensional spacetime, and the indices are field labels, like color or flavor indices. There should be no confusion between the two types of theories.} more commonly referred to as tensor models, were initially introduced in \cite{Ambjorn:1990ge,Sasakura:1990fs,Gross:1991hx} as a way to formulate quantum gravity as a sum over random geometries. With the same perspective, they were later revived by the discovery that they admit a new type of large-$N$ limit, dominated by melonic diagrams  \cite{Gurau:2010ba,Gurau:2011aq,Gurau:2011xq,Bonzom:2011zz}.\footnote{While in the original models were based on $r+1$ rank-$r$ tensors and a symmetry group such as $U(N)^{r(r+1)/2}$, the melonic large-$N$ limit has later been proved also for models with single rank-$r$ tensors and smaller symmetry groups, such as $U(N)^r$ \cite{Bonzom:2012hw}, $O(N)^r$ \cite{Carrozza:2015adg,Prakash:2019zia}, mixed products of orthogonal and unitary groups \cite{Tanasa:2015uhr}, and even irreducible tensor representations of a single group \cite{Klebanov:2017nlk,Benedetti:2017qxl,Carrozza:2018ewt}.}
However, it was only after a link to the Sachdev-Ye-Kitaev (SYK) model \cite{Sachdev:1992fk,Kitaev,Polchinski:2016xgd,Maldacena:2016hyu} was noticed \cite{Witten:2016iux,Klebanov:2016xxf} that the large-$N$ limit of tensor field theories in one or higher dimensions began being explored \cite{Peng:2016mxj,Krishnan:2016bvg,Bonzom:2017pqs,Beccaria:2017aqc,Gubser:2017qed,Giombi:2017dtl,Bulycheva:2017ilt,Choudhury:2017tax,Yoon:2017nig,Krishnan:2017lra,Prakash:2017hwq,Benedetti:2017fmp,Chang:2018sve,Giombi:2018qgp,Benedetti:2018ghn,Kim:2019upg,Benedetti:2019eyl,Klebanov:2019jup,Popov:2019nja,Benedetti:2019ikb,Benedetti:2019rja,Benedetti:2020yvb}.
For a comparison of the large-$N$ limits in  vector, matrix, and tensor field theories, see \cite{Klebanov:2018fzb}.

One reason of interest in the new melonic large-$N$ limit is that one typically finds a non-trivial conformal field theory (CFT) limit in the infrared, for which the full two-point function and the full spectrum of primary bilinear operators can be determined \cite{Klebanov:2016xxf,Giombi:2017dtl}.
Moreover, such \emph{melonic} CFTs can be obtained as non-trivial fixed points of the renormalization group, the latter being under control thanks to the use of the large-$N$ limit in combination with either an $\epsilon$-expansion away from the critical dimension \cite{Giombi:2017dtl} or with an expansion in an exactly marginal parameter \cite{Benedetti:2018ghn}, as we will review below.
Notice that a similar melonic truncation of the Feynman diagrams appeared already long time ago \cite{Patashinskii:1964,Gribov:1968fg}, as well as later in various other contexts, such as the so-called $\Phi$-derivable truncations studied for example in \cite{Blaizot:2003br,Berges:2005hc,Blaizot:2010zx}. However, the $1/N$ expansion offers a controlled way of implementing such approximations as a proper expansion scheme. 
Therefore, although so far no real-world application of such tensor field theories has emerged, they can play a useful role as toy models in improving our theoretical understanding of CFTs in higher dimensions, as well as possibly find theoretical applications in the AdS/CFT correspondence, or in connection to other theoretical models such as SYK.

The present survey aims at providing a short guide and introduction to recent results on melonic CFTs.
In Sec.~\ref{sec:general}, we give an introduction to the structure of the models, the graphical representation of invariants and of the perturbative expansion, and the dominant diagrams in the large-$N$ limit. In Sec.~\ref{sec:2PI}, we use the 2PI formalism \cite{Benedetti:2018goh} to introduce some essential features of tensor field theories, namely the Schwinger-Dyson equations for the two-point function, and the ladder structure of the four-point function.
Lastly, in Sec.~\ref{sec:boson} and \ref{sec:fermion}, we review the fixed points found in some bosonic and fermionic models, respectively.
We conclude in Sec.~\ref{sec:concl} with a summary and outlook.

%%%%%%%%%%%%%%%%%%%%%%%%
\section{Model building and melonic limit}
\label{sec:general}
%%%%%%%%%%%%%%%%%%%%%%%%

For simplicity, we will restrict our focus on models based on fields in the fundamental representation of $O(N)^3$ or $U(N)^3$,
that is, the fundamental field is a rank-3 field $\phi_{abc}(x)$ transforming as
\be \label{eq:field-trans}
\phi_{abc}(x) \to R^{(1)}_{aa'}\, R^{(2)}_{bb'}\,  R^{(3)}_{cc'}\,  \phi_{a'b'c'}(x)\, , %\;\;\;\; R^{(i)}\in \cG \,,
\ee
with $R^{(i)}$ being an element of $O(N)$ or $U(N)$ in the fundamental representation, and summation over repeated indices is implicit.
To fix ideas, we will also assume here that $\phi_{abc}(x)$ is a scalar field (real or complex, for the $O(N)$ or $U(N)$ case, respectively), while fermionic models will be discussed in Sec.~\ref{sec:fermion}.

As usual, a specific model is defined by the (Euclidean) spacetime dimension $d$, and a choice of action. We separate the latter in a free and an interacting part, as $S[\phi]=S_{\rm free}[\phi]+S_{\rm int}[\phi]$. The former is quadratic in the fields, thus defining the Gaussian part of the measure, with covariance $C(x,y)$:\footnote{Here and in the following we denote $\int_x=\int d^d x$ in direct space, and $\int_p=\int \f{d^d p}{(2\pi)^d}$ in momentum space. Repeated tensor indices imply a summation.}
\be
S_{\rm free}[\phi] = \int_{x,y} \phi_{abc}(x) C^{-1}(x,y)\phi_{abc}(y) \,.
\ee
For $d=0$ the standard choice is $C=1$, while for $d>0$ we choose the inverse covariance
\be \label{eq:free-C}
C^{-1}(x,y) = (-\p^2)^{\z} \equiv  \f{2^{2\z}\G\left(\f{d+2\z}{2}\right)}{\pi^{d/2}|\G(-\z)|} \f{1}{|x-y|^{d+2\z}}
\ee
endowing the field with the canonical dimension $\Delta_{\phi}= \f{d-2\zeta}{2}$. The choice of $\z$ distinguishes short-range models ($\zeta=1$)\footnote{The $\z\to1$ limit of the right-hand side of  \eqref{eq:free-C}  should be understood in the distributional sense, giving $C^{-1}(x,y) = -\p_x^2\d(x-y)$.} from long-range ones ($0<\zeta<1$).\footnote{Long-range models have a long history; in the scalar case with quartic interaction, they are known as the long-range Ising model \cite{Fisher:1972zz,Sak:1973}, which has been studied extensively with various methods, including constructive methods  \cite{Brydges:2002wq,Abdesselam:2006qg}, large-$N$ expansion \cite{Brezin:2014}, functional renormalization group \cite{Defenu:2014}, and CFT methods \cite{Paulos:2015jfa,Behan:2017emf}.}
One can of course also write directly the covariance by simply replacing $\z$ with $-\z$ in the above formula. Alternatively, we can use in momentum space the following formula, valid for $0<\z<1$:
\be
\tilde{C}(p) \equiv (p^2)^{-\z} =  \f{\sin(\pi\z)}{\pi}\int_0^{+\infty} ds\, \f{s^{-\z}}{p^2+s} \,.
\ee
This provides a K\"all\'en-Lehmann spectral representation of the propagator, showing that the spectral density $\f{\sin(\pi\z)}{\pi}s^{-\z}$ is positive  for $0<\z<1$.

For the interacting part we take a polynomial in the fields, without derivatives, restricted by invariance under  \eqref{eq:field-trans}.
The invariance requires that tensor indices in the same position on different tensors be contracted pairwise (implying in particular that invariants are monomials of even order).
For example, the so-called \emph{tetrahedron} invariant is defined as
\be \label{eq:I-tetr}
I_{\rm tetrahedron} = \phi_{a_1 a_2 a_3} \phi_{a_1 b_2 b_3}  \phi_{b_1 a_2 b_3} \phi_{b_1 b_2 a_3} \,.
\ee
Given a set of invariants of this sort, we write the interacting part of the action as
\be \label{eq:gen-int}
S_{\rm int}[\phi] = \int_x  \sum_q \sum_{b \in \cI_q} \f{\l_b}{2q\, N^{\f32(q-1)+\r_{b}}} I_{b} 
=  \int_x  \sum_q \sum_{b\in \cI_q} \f{\l_b}{2q\, N^{\f32(q-1)+\r_{b}}} \d^{(b)}_{\mba_1 \ldots \mba_{2q}}\phi_{\mba_1}(x)\ldots \phi_{\mba_{2q}}(x)\,.
\ee
Here, $\cI_q$ is a set of labels that distinguishes the different invariants built out of $2q$ fields, and we have introduced a condensed notation for the indices, $\mba_i=a_{i1}a_{i2}a_{i3}$, and an invariant tensor of rank $6q$, $\d^{(b)}_{\mba_1 \ldots \mba_{2q}}$, which performs the contraction of tensor indices corresponding to a given invariant.
The $\l_b$ are `t Hooft couplings (to be held fixed in the large-$N$ limit), and the scaling in $N$, in particular the parameters $\r_b$, has to be chosen in such a way that the large-$N$ limit exists and it is non-trivial. We will get back to it after having introduced a graphical representation.

\begin{figure}[htb]
 \begin{center}
 \includegraphics[width=9cm]{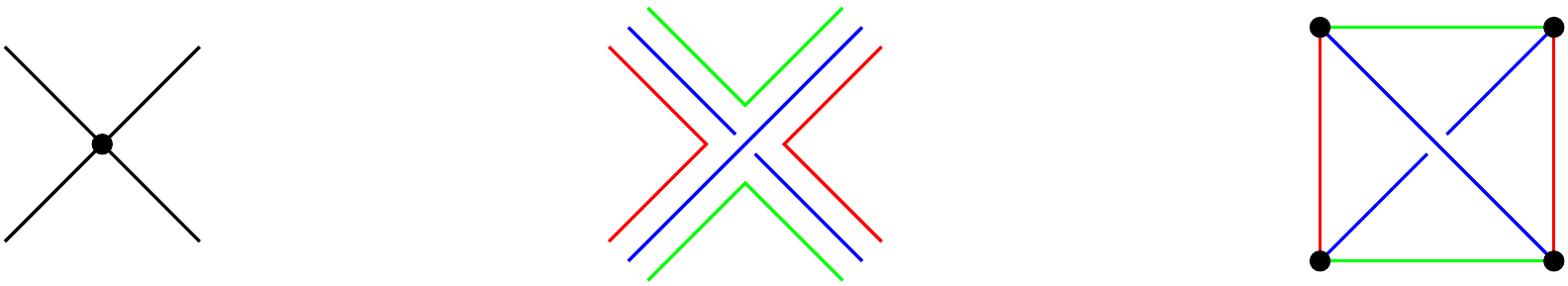}  
 \caption{The tetrahedral vertex in three different representations; from left to right: the Feynman diagram representation, the stranded representation, and the 3-colored graph representation. In the last two, the colors label the different indices of a tensor.} \label{fig:tensorVertex}
 \end{center}
 \end{figure}

In the Feynman diagram representation, a monomial of order $2q$ in the field is represented by a $2q$-valent vertex; while this is a useful representation for the spacetime integrals arising from Wick contractions because all the fields in a monomial are at the same spacetime point, it clearly misses the tensor structure, and it cannot help us in distinguishing different invariants built out of the same number of fields. To take that into account, two types of representations are typically employed, in which the perturbative expansion is represented by either stranded graphs or edge-colored graphs. The former is a natural generalization of the ribbon graphs of matrix models, but it can result in rather heavy drawings, as each propagator is represented by three (or more, for higher rank models) parallel lines.
It is then more convenient to represent the tensor invariants as {\it colored graphs} \cite{RTM}: we represent every tensor field as a node (black and white for $\phi$ and $\phib$, respectively, if the field is complex) and every contraction of two indices as an edge. Each edge is assigned a color red, blue, or green (or a label $1$, $2$, or $3$) corresponding to the positions of the indices in the tensor. We call the resulting graphs $3$-colored graphs (or sometimes \emph{bubbles}). The three different representations of the tetrahedron invariant \eqref{eq:I-tetr} are depicted in Fig.~\ref{fig:tensorVertex}.

\begin{figure}[ht]
\begin{center}
\includegraphics[width=0.6\textwidth]{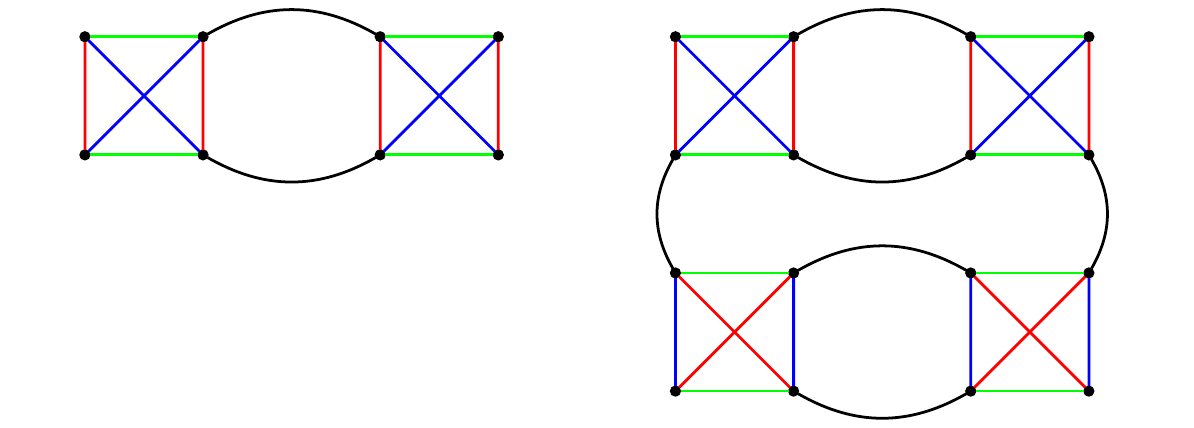}
 \caption{Two $4$-colored graphs in the perturbative expansion of the four-point function, with tetrahedron interaction vertices drawn in red, blue, and green colors, and propagators drawn in black.
 The resulting structure of index contractions of the external tensor are equivalent to the pillow (left) and double-trace (right) invariants.} \label{fig:graph}
 \end{center}
\end{figure}

\begin{figure}[ht]
\begin{center}
\includegraphics[width=0.6\textwidth]{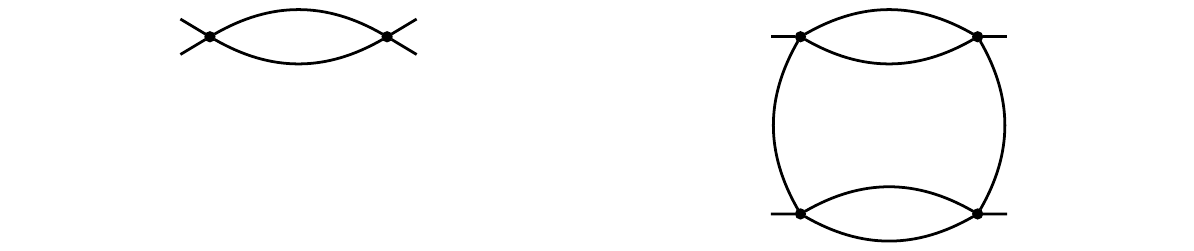}
 \caption{Two Feynman diagrams obtained from the two $4$-colored graphs of Fig.~\ref{fig:graph} by shrinking the colored edges. Half-edges are also added to keep track of the external fields.} \label{fig:FeynDiagr}
 \end{center}
\end{figure}

In the perturbative expansion around the free theory, we represent the free propagators as edges of a new color, connecting two different nodes (the two tensor fields whose Wick contraction leads to that propagator). We choose the black color for the propagator lines, or equivalently, the label $0$. When representing the interaction bubbles as 3-colored graphs, the perturbative expansion is then captured by \emph{$4$-colored graphs}. We give two examples of $4$-colored graphs in Fig.~\ref{fig:graph}, to be compared with the respective standard Feynman diagrams in Fig.~\ref{fig:FeynDiagr}.
The latter offer a more intuitive  representation of Feynman integrals, but the $4$-colored graphs are necessary in order to identify the scaling in $N$. 
Indeed, in a $4$-colored graph, each propagator identifies all three indices on its two end tensors, whereas each edge of color $i$ identifies only one pair of indices between its end tensors. The indices will then circulate along cycles of color $0i$ with fixed $i$, which we call \emph{faces}, hence each face gives rise to a free sum, that is, a factor $N$. 
The amplitude of a Feynman diagram $\mathcal{G}$ thus scales as $A(\mathcal{G})\sim N^{F-\sum_b \r_b n_b}$, with $F$ the total number of faces in the associated $4$-colored graph and $n_b$ the number of bubbles of the interaction $b$. 
The existence of the large-$N$ limit relies on the fact that the power of $N$ is bounded from above for an appropriate choice of $\r_b$ \cite{RTM,Carrozza:2015adg}. Following \cite{Carrozza:2015adg}, we take
\be \label{eq:opt-scal}
\r_b=\frac{F(I_b)-3}{2} \,,
\ee
with $F(I_b)$ counting the total number of cycles of alternating colors $i$ and $j$ with $i,j ~\in \lbrace 1,2,3\rbrace$ in the 3-colored graph representing the invariant $I_b$.

\paragraph{Melonic graphs and melonic diagrams.} 

\begin{figure}[htb]
\begin{center}
\includegraphics[width=0.2\textwidth]{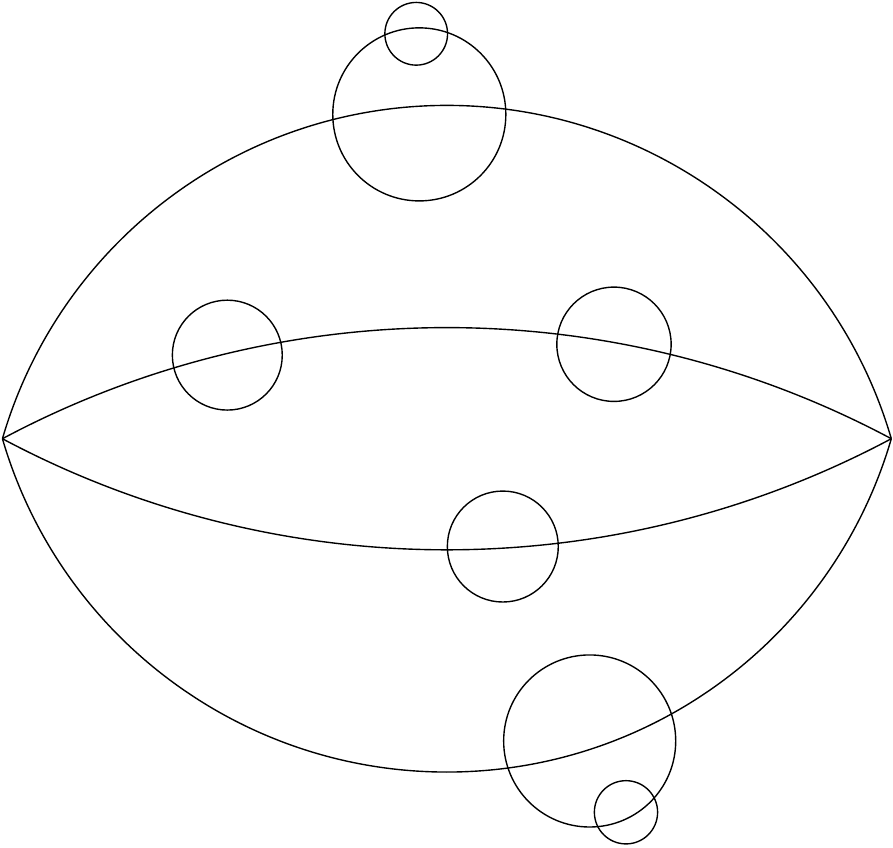}
 \caption{A melonic 4-valent graph.} \label{fig:melon}
\end{center}
\end{figure}

Melonic $k$-valent graphs are defined constructively starting from the fundamental melon, i.e.\ the unique graph built out of two $k$-valent vertices without forming self-loops (or tadpoles), and then iteratively inserting on any edge a melonic 2-point function, i.e.\ the graph obtained from the fundamental melon by cutting one edge in the middle (also known as sunset diagram). An example is given in Fig.~\ref{fig:melon}. Notice that melonic $k$-valent graphs are always bipartite, and edge-colorable with $k$ colors. 
%%%%%%%
%%%%%%%

An important result in rank-$r$ tensor models is that if one only allows for interaction bubbles which are melonic $r$-valent graphs, then in the perturbative expansion the leading order vacuum graphs at large $N$ are melonic $(r+1)$-valent graphs \cite{Bonzom:2012hw}.
However, it is important to notice that melonic $(r+1)$-valent graphs do not correspond to melonic Feynman diagrams, i.e.\  they do not remain melonic after shrinking the colors from 1 to $r$.
From the point of view of the Feynman diagrams, melonic $(r+1)$-valent graphs reduce to the same type of cactus diagrams appearing in the large-$N$ limit of vector models, and therefore field theories based on such interaction are not expected to lead to very different results than vector models.\footnote{They can nevertheless lead to new phases with patterns of spontaneous symmetry breaking which are impossible in the vector case \cite{Benedetti:2018ghn}.} 

Adding non-melonic bubbles, things get more complicated, and possibly more interesting. In particular, it was found in Ref.~\cite{Carrozza:2015adg} that non-melonic interaction bubbles, such as  the tetrahedron invariant \eqref{eq:I-tetr}, can be scaled in such a way that they also contribute at leading order in the $1/N$ expansion, and that for some interactions (in that specific example, the quartic tetrahedron interaction) their leading-order Feynman diagrams are melonic. The possibility of restricting the spacetime Feynman diagrams to the melonic type by means of a large-$N$ limit has been a main reason for studying tensor field theories in dimension $d\geq 1$, starting from \cite{Klebanov:2016xxf}.
In this respect, we mention a few interesting studies: for interactions of higher order, Ref.~\cite{Lionni:2017yvi} identified the structure of diagrams at first leading orders;
Ref.~\cite{Ferrari:2017jgw} proved that, with tensors of prime rank and for a particular class of complete interactions (i.e.\ invariants corresponding to a complete graph), the dominant Feynman diagrams are melonic;
Ref.~\cite{Gubser:2018yec} classified the different structures of complete interactions for tensors of odd rank;
Ref.~\cite{Prakash:2019zia} showed melonic dominance in sextic subchromatic models (rank-3 and a particular rank-4 model).

For concreteness, we concentrate here on interactions which are at most quartic,\footnote{Notice also that the number of possible interactions at higher orders grows very fast; for an explicit counting, see \cite{Geloun:2013kta,BenGeloun:2017vwn,Avohou:2019qrl}, as well as \cite{Bulycheva:2017ilt}.} resulting in the most studied version of the potential:
\be \label{eq:int-action-graph}
\begin{split}
S_{\rm int}[\phi] = & \int_x  \left(\f12 m^{2\z} \,\vcenter{\hbox{\includegraphics[width=1.6cm]{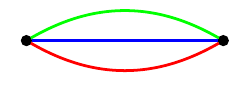}}}
+ \f{\l_t}{4 N^{3/2}} \,\vcenter{\hbox{\includegraphics[width=1.6cm]{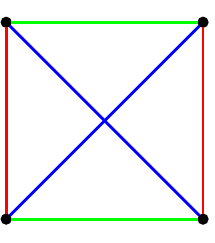}}}  \right.\\
&\qquad\left.  + \f{\l_p}{12 N^2}  \sum_{\text{col.perm.}} \vcenter{\hbox{\includegraphics[width=1.6cm]{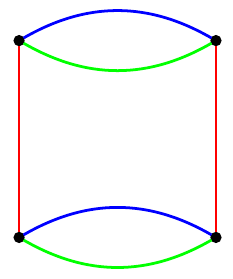}}}  
 +\f{\l_d}{4 N^3} \, \vcenter{\hbox{\includegraphics[width=1.6cm]{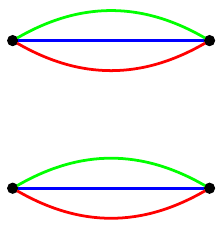}}}\right)\,.
\end{split}
\ee
Here, the first invariant is a mass term, while the second is the tetrahedron, which we have already encountered, and which is the only complete interaction of order four; lastly, the third and fourth invariants are known as pillow and double-trace, respectively. Notice that there are actually three pillow invariants, distinguished by the choice of color on the vertical edges: by summing over color permutations (which we have divided by a factor 3 for convenience), with the same coupling $\l_p$, we enforce a color symmetry on the action.
In general, bubbles which are composed of one or several connected components are referred to as single-trace or multi-trace, respectively, for analogy with the matrix case, and bubbles $I_b$ for which $\rho_b=0$ are called \emph{maximally single trace}  (MST), as each of their 2-colored subgraphs are connected, i.e.\ they are like the graph of a matrix single-trace invariant.
The mass and tetrahedron invariants are the only MST bubbles in our action.

\paragraph{The $1/N$ expansion.}
The model with interaction \eqref{eq:int-action-graph} has a $1/N$ expansion \cite{Carrozza:2015adg,Klebanov:2016xxf}. The simplest way to see this is to observe that pillow and double-trace vertices can be obtained as radiative corrections from the tetrahedral vertex: the pillow is a rung (Fig~\ref{fig:graph}, left), and the double-trace 
is a ladder made out of two rungs with different color inside their loop (Fig~\ref{fig:graph}, right). Replacing the pillow and double-trace vertices in a graph by their minimal resolution in terms of tetrahedral vertices one associates with any graph $\cG$ 
a graph $\hat \cG$ having \emph{only} tetrahedral vertices but the same scaling in $N$:
\be
 F(\cG) -\frac{3}{2}n_t(\cG) - 2 n_{p}(\cG) - 3n_{d}(\cG) =  F(\hat \cG) -\frac{3}{2}n_t( \hat \cG) \,.
\ee
Starting from $\hat \cG$ one can build three \emph{jackets} \cite{Gurau:2010ba,Carrozza:2015adg} ${\cal J}^i$, that is ribbon graphs\footnote{The ribbon graphs are made evident in the stranded representation, where one replaces each black line and vertex by three parallel red, green, and blue lines: a jacket ${\cal J}^i$ is then obtained by simply deleting color $i$.} obtained by ignoring the faces of color $0i$.
Each jacket has a non orientable genus $k({\cal J}^i) \ge 0$ and the number of faces\footnote{It is at this point that one uses the fact that $\hat \cG$ has only tetrahedral vertices. This construction is slightly more complicated on the original graph $\cG$, as the jackets of $\cG$ are not necessarily connected \cite{Carrozza:2015adg}.}
$ {\cal F}(\cJ^i) =  n_t(\hat \cG) + 2 - k({\cal J}^i)$.
As every face belongs to two jackets, the total number of faces of $\hat \cG$ is
\be
 {\cal F}(\hat \cG) = \frac{3}{2} n_t(\hat \cG) + 3 - \frac{1}{2} \sum_{i} k({\cal J}^i) \; .
\ee
Denoting $ \omega(\cG) = \frac{1}{2} \sum_{i} k({\cal J}^i) \ge 0$ the \emph{degree} of the original graph $\cG$, the scaling with $N$ of a connected vacuum graph is
\be \label{eq:N-graph-amplitude}
 N^{3 - \omega(\cG)} \; .
\ee

\begin{figure}[ht]
\begin{center}
\includegraphics[width=0.3\textwidth]{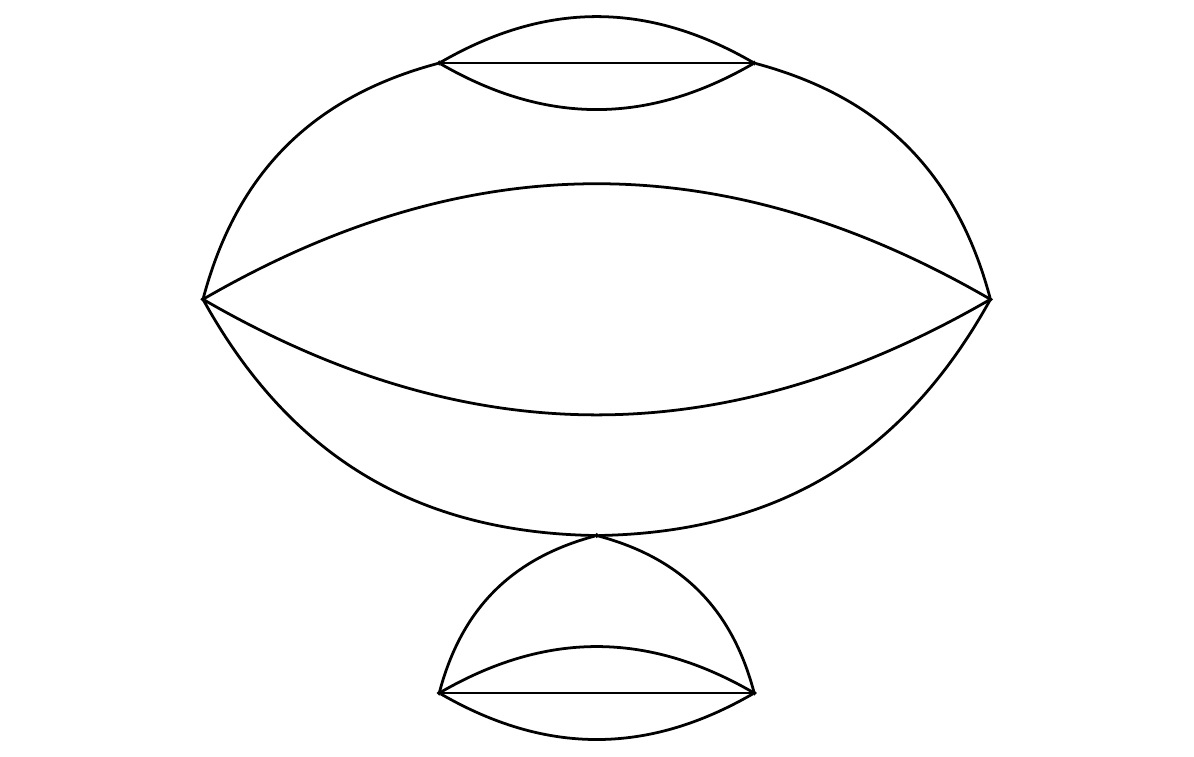}
 \caption{A melon-tadpole diagram, where all the invariants have been shrunk to point-like vertices.} \label{fig:melontadpoles}
 \end{center}
\end{figure}

By the standard arguments \cite{Bonzom:2011zz,RTM} $\cG$ has degree zero if and only if $\hat \cG$ is melonic.
That is the leading order graphs are melonic \emph{after} substituting all the pillows and double-trace vertices by their minimal realizations in terms of 
the tetrahedral vertex. In terms of the original interactions in $\cG$, one gets \emph{melon-tadpole} \cite{Benedetti:2017qxl} graphs, that is, 
graphs obtained by iterated insertions of melons or tadpoles into melons or tadpoles,
see Fig.~\ref{fig:melontadpoles}. Observe that all the tadpoles are based on
either pillow or double-trace vertices, while the end vertices of the melons are tetrahedral.

%%%%%%%%%%%%%%%%%%%%%%%%
\section{2PI formalism, Schwinger-Dyson equations, and four-point kernel}
\label{sec:2PI}
%%%%%%%%%%%%%%%%%%%%%%%%

The melonic limit allows us to write in closed form the Schwinger-Dyson (SD) equations for the two- and four-point functions, that otherwise are usually known only perturbatively.
Such equations can be conveniently formulated in terms of a generating functional. However, from figures \ref{fig:melon} and \ref{fig:melontadpoles} it is clear that we still have an infinite series of diagrams, hence it is not obvious how to write the generating functionals of connected or one-particle irreducible (1PI) diagrams. On the other hand, it is clear that all the melon-tadpoles diagrams are obtained from some fundamental diagram by repeated insertions of two-point functions corrections. This suggests to consider the two-particle irreducible (2PI) effective action, in whose perturbative expansion no two-point insertions are allowed \cite{Cornwall:1974vz} (hence the name: a 2PI diagram is a diagram that cannot be disconnected by cutting two propagators). In such formalism, the Feynman diagrams are 2PI, but have an unknown propagator $G$, which is to be fixed a posteriori from the field equations of the 2PI effective action, which coincide with the SD equation for the two-point function, as we review in subsection~\ref{sec:SDeq}.
%Similarly, in subsection~\ref{sec:BSeq}, we will see how the BS equations arise from the second functional derivative of the  2PI effective action.
 
We will now briefly review the essential features of the 2PI effective action. For a general introduction to the formalism, see the original \cite{Cornwall:1974vz} or the review \cite{Berges:2004yj}; for its application to tensor models, and a more detailed version of our presentation, see \cite{Benedetti:2018goh,Benedetti:2019eyl,Gurau:2019qag}.
We will assume that there is no spontaneous symmetry breaking; for a discussion of how the formalism could be applied to a broken phase, see \cite{Benedetti:2019sop}.

The 2PI effective action is obtained by introducing a bilocal source in the functional integral, coupled to the bilinear $\f12 \phi_{\mba}(x)\phi_{\mbb}(y)$, and performing a Legendre transform of the logarithm of the partition function with respect to the source.
As a result of such procedure, the full 2PI effective action can be written as
\be \label{eq:2PI-def}
\G [ G] =  \frac{1}{2} \Tr\left[C^{-1}  G\right] + \f12 \Tr[\ln G^{-1}] +\G_2 [G] \,,
\ee
where
\be \label{eq:Gamma_2}
\G_2 [ G] = -\ln \int_{2PI} d\m_{G}[\vph] \; e^{  -  {S}_{\rm int}[\varphi] } \,.
\ee
In the functional integral, $d\m_{G}[\vph]$ is a normalized Gaussian measure with covariance $G_{abc,a'b'c'}$, and the subscript 2PI reminds us that in the perturbative expansion we only retain 2PI diagrams.
Traces and products of bilocal functions are in the matrix sense, with a matrix index collectively corresponding to an $O(N)^3$ triplet of indices and a spacetime point.
For example,
\be
\Tr\left[C^{-1} G\right] \equiv  \int_{x,y} (C^{-1})_{\mba,\mba'}(x,y) G_{\mba',\mba}(y,x) = \int_{x,y} C^{-1}(x,y) G_{\mba,\mba}(y,x)\,,
\ee
where in the last step we used the fact that our free covariance is diagonal in the tensor indices.
The two traces in \eqref{eq:2PI-def} represent the classical and one-loop parts of the effective action, while $\G_2[G]$ contains all the higher-loops contributions.

The full two-point function is to be determined by the field equation:
\be \label{eq:SDeq-gen}
\f{\d \G [ G]}{\d G_{\mba,\mbb}(x,y) } = 0 \;\;\; \Leftrightarrow  \;\;\; (G^{-1})_{\mba,\mbb}(x,y) = C^{-1}(x,y) \d_{\mba,\mbb} +2 \f{\d \G_2 [ G]}{\d G_{\mba,\mbb}(x,y) } \,,
\ee
where $\d_{\mba,\mbb}=\d_{a_1b_1}\d_{a_2b_2}\d_{a_3b_3}$.
The field equations have exactly the form of the SD equation, under the identification of the self-energy
\be
\Si_{\mba,\mbb}(x,y) = -2 \f{\d \G_2 [ G]}{\d G_{\mba,\mbb}(x,y) }\,.
\ee
Since cutting one propagator from 2PI diagrams (i.e.\ deriving with respect to $G$) leads to 1PI diagrams, we obtain that the self-energy is a sum of 1PI diagrams, as expected.

\begin{figure}[htb]
\begin{center}
\includegraphics[width=0.6\textwidth]{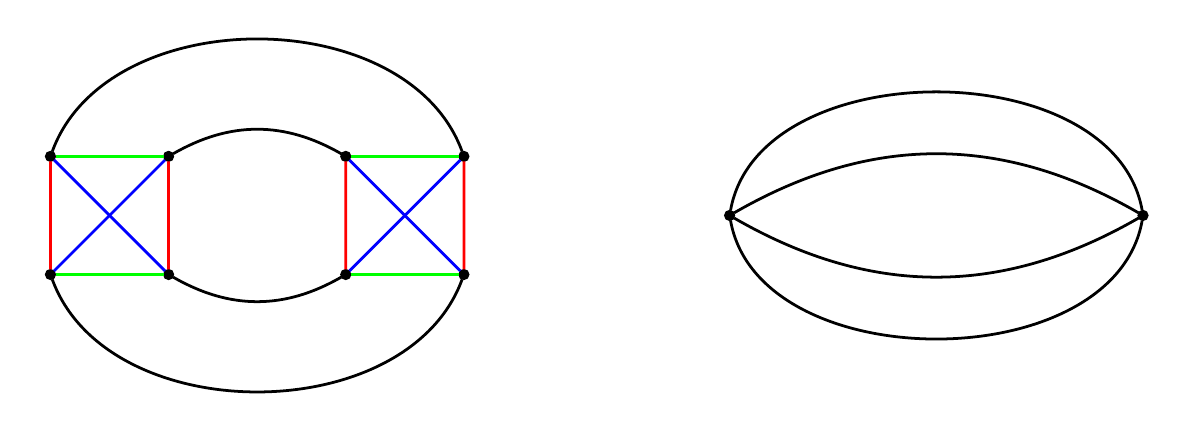}
 \caption{A vacuum 4-colored graph (left) built on two tetrahedron interactions, corresponding to a melonic  Feynman diagram (right).} \label{fig:vacuum-2PI-melon}
 \end{center}
\end{figure}

\begin{figure}[htb]
\begin{center}
\hspace{-1cm}\includegraphics[width=0.5\textwidth]{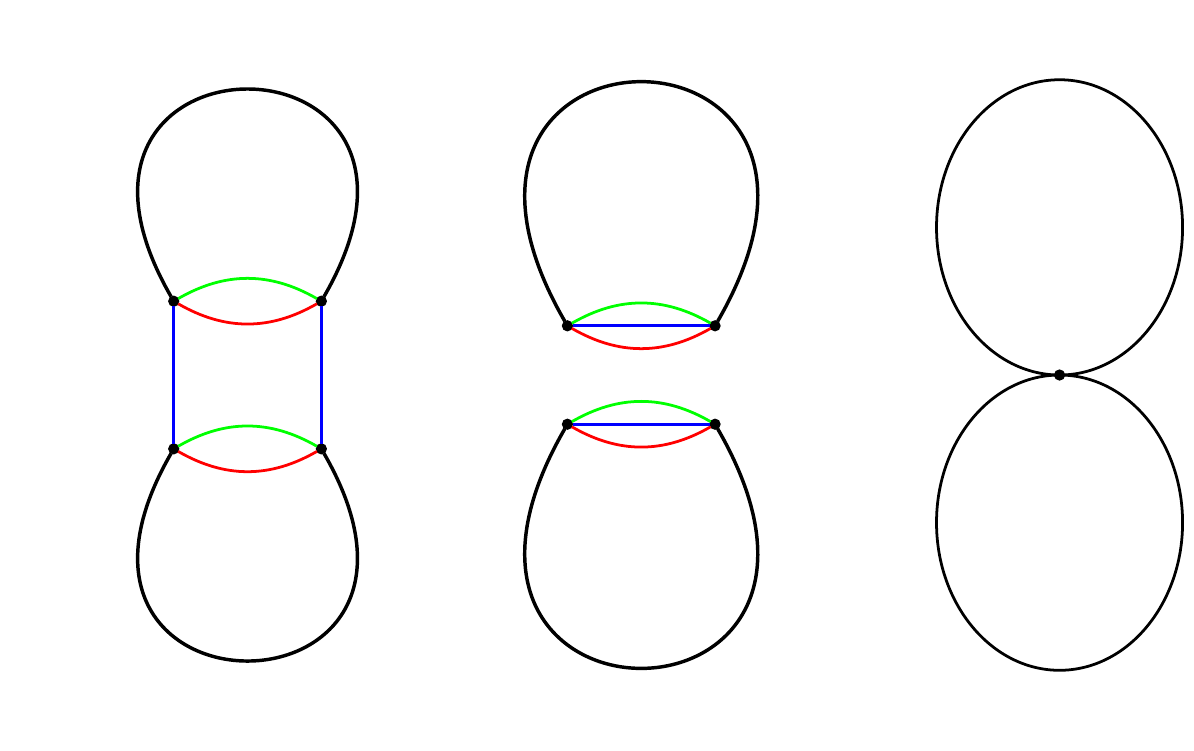}
 \caption{The two type of vacuum 4-colored graphs (left and center) corresponding to the figure-eight Feynam diagram (right), occurring at leading order in $1/N$.} \label{fig:vacuum-2PI-8}
 \end{center}
\end{figure}

In the symmetric phase, the full two-point function is diagonal in the tensor indices:
\be \label{eq:G-id}
G_{\mba,\mbb}(x,y) = G(x,y)\, \d_{\mba,\mbb}\,,
\ee
hence $\Tr\left[C^{-1} G\right]  = N^3 \int_{x,y} (C^{-1})(x,y) G(y,x)$. Similarly, the one-loop trace in \eqref{eq:2PI-def} will also be of order $N^3$. Furthermore, as the propagators are proportional to the identity, as in the original theory, the counting of factors of $N$ is unchanged, hence  the result \eqref{eq:N-graph-amplitude} is still valid.
Therefore, for any interaction action as in \eqref {eq:gen-int}, with $\r_b$ as in \eqref{eq:opt-scal}, we have the following expansion:
\be \label{eq:KTCT-expansion}
\G_2[0,G] = \sum_{\omega\in \mathbb{N}/2} {\G}^{ ( 3-\omega ) }_2[G]   \,,\qquad \text{with}\;\; {\G}^{ (p) }_2[G] \sim N^p \,.
\ee
For the particular choice of action \eqref{eq:int-action-graph}, at leading order we have melon-tadpoles diagrams. Imposing on such diagrams also the 2PI condition drastically reduces the number of leading order diagrams to just five: a melon with two tetrahedron bubbles (see Fig.~\ref{fig:vacuum-2PI-melon}), and four figure-eight diagrams, three with one of the pillow bubbles and one with the double-trace (Fig.~\ref{fig:vacuum-2PI-8}).
In addition, if we do not include a mass term in the free part of the action, as we chose to do, we also need to add the one-loop diagram with one two-valent vertex.
As a result, the 2PI effective action at leading order in $1/N$ is
\be \label{eq:2PI-result}
\begin{split}
\G_2 [G] = & \f12 m^{2\z} \int_x G_{a_1 a_2 a_3,a_1 a_2 a_3}(x,x)+ % \\&
 \f{\l_d}{4N^3} \int_x \left(G_{a_1 a_2 a_3,a_1 a_2 a_3}(x,x)\right)^2 \\
&+\f{\l_p}{12N^2} \sum_{i=1}^3  \d_{a_ia'_i}\d_{b_ib'_i} (\prod_{j\neq i} \d_{a_jb_j} \d_{a'_jb'_j})\int_x G_{a_1 a_2 a_3,b_1 b_2 b_3}(x,x) G_{a'_1 a'_2 a'_3, b'_1 b'_2 b'_3}(x,x)\\
&-\f{\l_t^2}{8 N^3} \int_{x,y}G_{a_1 a_2 a_3, b_1 b_2 b_3}(x,y) G_{a_1 a'_2 a'_3, b_1 b'_2 b'_3}(x,y)  G_{a'_1 a_2 a'_3, b'_1 b_2 b'_3}(x,y) G_{a'_1 a'_2 a_3, b'_1 b'_2 b_3}(x,y) \,.
\end{split}
\ee
Notice that upon substitution of $G_{\mba,\mbb}$ by \eqref{eq:G-id}, we find that all terms are indeed proportional to $N^3$.
Strikingly, at leading order in $1/N$ we can write the full 2PI in closed form. The closed expression \eqref{eq:2PI-result} is another manifestation of the fact that the melonic limit of tensor models lies somewhere in between the planar limit of matrix models, for which a closed expression for the effective action is generally not possible, and the cactus limit of vector models, for which a closed expression is possible \cite{Aarts:2002dj,Berges:2001fi}, but with a less interesting ultralocal structure at leading-order, identical to the double-trace- contribution in the first line of \eqref{eq:2PI-result}. The melon contribution in the last line of \eqref{eq:2PI-result} is bilocal, thus leading to different physics than the vector model. This is an important feature of the melonic contributions, and it is also at the heart of the most interesting aspects of the SYK model.

%%%%%%%%%%%%%%%%%%%%%%%%
\subsection{Melonic SD equation}
\label{sec:SDeq}
%%%%%%%%%%%%%%%%%%%%%%%%

\begin{figure}[htb]
\begin{center}
\includegraphics[width=0.9\textwidth]{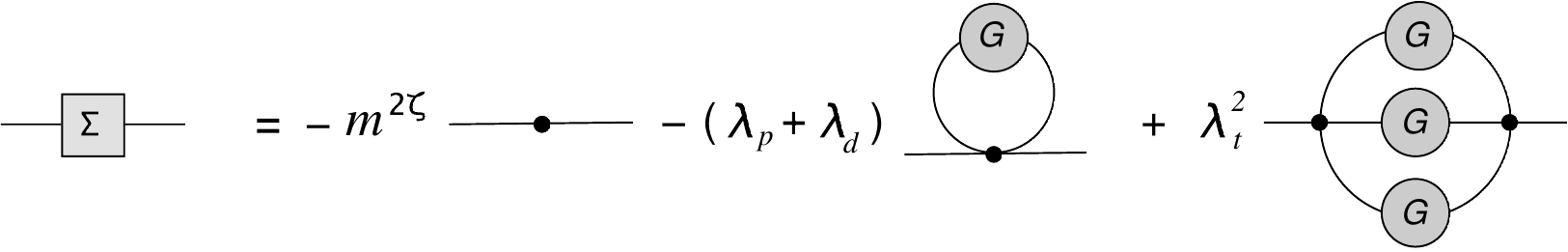}
 \caption{The self-energy $\Sigma(x,y)$ in Eq.~\eqref{eq:SDE-LO}.} \label{fig:full-quartic-Sigma}
 \end{center}
\end{figure}

By plugging \eqref{eq:2PI-result} into \eqref{eq:SDeq-gen}, and using \eqref{eq:G-id}, we find the melonic SD equation:
\be \label{eq:SDE-LO}
\begin{split}
G^{-1}(x,y) & = C^{-1}(x,y) - \Sigma(x,y) \\
&=C^{-1}(x,y) + \d(x-y) \left( m^{2\z} +(\l_d +\l_p) G(x,x)\right) -\l_t^2 G(x,y)^3 \,.
\end{split}
\ee
The self energy $\Sigma(x,y)$ has the graphical interpretation in Fig.~\ref{fig:full-quartic-Sigma}.

In momentum space this becomes:
\be\label{eq:SDE-p}
\begin{split}
%  \Sigma(p) & = - m^{2\zeta}    +     \lambda^2    \int_{q_1,q_2}    G(q_1)  G(q_2)  G( p +q_1 + q_2  )    - ( \lambda_p + \lambda_d ) \int_{q}    G(q) \,, \crcr
   G(p)^{-1} %&
     =   C(p)^{-1} + m^{2\zeta} +( \lambda_p + \lambda_d ) \int_{q}    G(q) - \lambda_t^2    \int_{q_1,q_2}    G(q_1)  G(q_2)  G( p +q_1 + q_2  ) \; .
\end{split}
\ee 
The bare mass can be tuned to cancel with the tadpoles and with the melon contribution at $p=0$.
Then, for $C(p)^{-1}=p^2$ (i.e.\ for $\zeta=1$), a simple power counting argument indicates that the solution admits two regimes \cite{Klebanov:2016xxf}: a free scaling regime in the ultraviolet $G(p)^{-1}\sim p^2$ (with $C(p)^{-1}$ dominating over the self-energy), and an anomalous scaling regime in the infrared $G(p)^{-1}\sim p^{d/2}$ (with the self-energy dominating over $C(p)^{-1}$).
Choosing instead, as in  \cite{Benedetti:2019eyl}, $\zeta = d/4$ to match the infrared conformal behavior, one finds a scaling behavior for all $p$.
In fact, with $C(p)^{-1}=p^{2\zeta}$, the Schwinger-Dyson equation is formally solved by $ G(p)^{-1} = Z p^{2\zeta}$, as can be naively seen by plugging it into \eqref{eq:SDE-p},
\be\label{eq:formal}
  Z p^{2\zeta} = p^{2\zeta} +  m^{2\zeta} - \frac{\lambda_t^2}{Z^3}    \int_{q_1,q_2}   \;\frac{1}{q_1^{2\zeta}} \; \frac{1}{q_2^{2\zeta}} \; \frac{1}{ ( p +q_1 + q_2  )^{2\zeta} } +  \frac{  \lambda_p + \lambda_d }{Z}   \int_{q} \;\frac{1}{ q^{2\zeta} }  \; ,
\ee 
and noticing that the double integral (which we call the  melon integral) gives, after a rescaling of $q_1$ and $q_2$ by $|p|$, a global $|p|^{2d - 6\zeta} = |p|^{2\zeta}$. Differently from the $\z=1$ case of \cite{Klebanov:2016xxf}, here there is only one regime: $\Sigma(p)$ and $C(p)^{-1}$ are of the same order in $p$.
However, both integrals in Eq.~\eqref{eq:formal} are divergent, thus we need regularization and renormalization. The full procedure is detailed in \cite{Benedetti:2019eyl}, and we will not repeat it here.
The final result of that analysis is that, for  $\zeta= \frac{d}{4}$, choosing $m^{2\zeta}$ to cancel the tadpoles and the $p=0$ part of the melon integral, the SD equation \eqref{eq:SDE-p} is solved by $G(p)^{-1}=Zp^{2\z}$, with $Z$ satisfying:
\be\label{eq:wave}
Z^4 - Z^3  =   \lambda_t^2 \frac{1}{(4\pi)^d } \;\frac{\Gamma \left( 1 -\frac{d}{4} \right) }{ \frac{d}{4}\Gamma\left( 3 \frac{d}{4}\right)} \,.
\ee
It should be stressed that $Z$ is finite for $\z=d/4$ in $d<4$, hence it is not a wave function renormalization, but rather a function that resums all the melonic insertions in the propagator.

%%%%%%%%%%%%%%%%%%%%%%%%
\subsection{Melonic four-point kernel} 
\label{sec:4pt-K}
%%%%%%%%%%%%%%%%%%%%%%%%

Another important equation satisfied by the 2PI effective action concerns its second derivative with respect to $G$:
\begin{equation} \label{eq:Hess-inv}
 \int_{u,v}\cF_{ (\mba , \mbb ) ; ( \mbc , \mbd )}(x,y,u,v) \f{\d^2 \G}{\d G_{\mbc  \mbd}(u,v)\d G_{\mbm \mbn}(w,z)}   = \mathbb{I}_{\mba \mbb ; \mbm \mbn}(x,y,w,z) \,,
\end{equation}
where we have introduced the projector on symmetric bilocal matrices
\begin{equation}  \label{eq:symm-proj}
\mathbb{I}_{\mba \mbb ; \mbm \mbn}(x,y,w,z) = \frac{1}{2} (\delta_{\mba \mbm} \delta_{\mbb \mbn} \d(x-w)\d(y-z)+\delta_{\mba \mbn} \delta_{\mbb \mbm}\d(x-z)\d(y-w))\,,
\end{equation}
and the forward four-point function
\be \label{eq:4pt-fwd}
\begin{split}
\cF_{ (\mba , \mbb ) ; ( \mbc , \mbd )}(x,y,w,z) 
= & \Braket{ \phi_\mba(x) \phi_\mbb(y) \phi_\mbc(w) \phi_\mbd(z)} - \Braket{ \phi_\mba(x) \phi_\mbb(y)} \Braket{ \phi_\mbc(w) \phi_\mbd(z)} \\
= & \Braket{ \phi_\mba(x) \phi_\mbb(y) \phi_\mbc(w) \phi_\mbd(z)}_c  \\
&+  \Braket{ \phi_\mba(x) \phi_\mbc(w) }\Braket{\phi_\mbb(y)  \phi_\mbd(z)} +\Braket{ \phi_\mba(x) \phi_\mbd(z) }\Braket{\phi_\mbb(y)  \phi_\mbc(w)} \,.
\end{split}
\ee
Notice that if the expectation values here are evaluated without source term in the functional integral, then \eqref{eq:Hess-inv} holds for $G$ on shell, i.e.\ for $G$ that solves the SD equation.

\begin{figure}[ht]
\begin{center}
\includegraphics[width=0.65\textwidth]{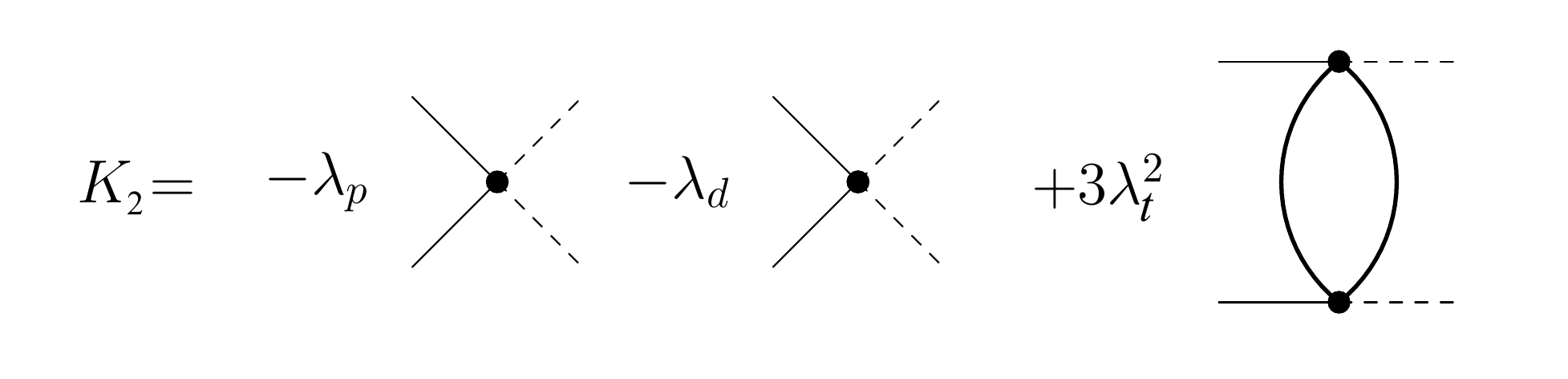}
 \caption{Graphical representation of the kernel \eqref{eq:K2}. Solid lines represent full two-point functions, while dashed lines represent amputated external legs. The first two terms are based respectively on pillow and double-trace vertices while the last one is based on a pair of tetrahedral vertices.} \label{fig:kernel}
 \end{center}
\end{figure}

On the other hand, using \eqref{eq:2PI-def}, we find:
\be \label{eq:Hess-K}
\f{\d^2 \G}{\d G_{\mba  \mbb}(x,y)\d G_{\mbm \mbn}(w,z)}
   = \f12 \int_{u,v} G^{-1}_{\mba  \mbc}(x,u)G^{-1}_{\mbb \mbd}(y,v) \left(\mathbb{I}-K\right)_{\mbc \mbd ; \mbm \mbn}(u,v,w,z) \,,
\ee
where we introduced the four-point kernel
\be \label{eq:K-def}
K_{\mba \mbb ; \mbm \mbn}(x,y,w,z) =\int_{u,v}  G_{\mba  \mbc}(x,u)G_{\mbb \mbd}(y,v) \f{\d \Si_{\mbm,\mbn}(w,z) }{\d G_{\mbc \mbd}(u,v)} \,.
\ee
Combining \eqref{eq:Hess-inv} and \eqref{eq:Hess-K}, we obtain:
\be \label{eq:4pt-fwd-ladder}
\cF_{ (\mba , \mbb ) ; ( \mbm , \mbn )}(x,y,w,z) 
 = \int_{u,v} \left(\mathbb{I}-K\right)^{-1}_{\mba \mbb ; \mbc \mbd}(x,y,u,v) \left(G_{\mbc  \mbm}(u,w)G_{\mbd \mbn}(v,z) + G_{\mbc  \mbn}(u,z)G_{\mbd \mbm}(v,w) \right)\,.
\ee
This is a general equation, as we have not made any approximation so far, and as such it is rather formal.
However, in the large-$N$ limit, the 2PI effective action is given explicitly from \eqref{eq:2PI-def} and \eqref{eq:2PI-result}, hence we have an explicit expression for the four-point kernel. The inversion of the tensor structure in $\mathbb{I}-K$ can also be performed explicitly, after recognizing that $K$ can be decomposed in a sum of orthogonal projectors; we refer to \cite{Benedetti:2019eyl} for such details.
Here we just provide the expression for the double trace of $\cF$, which is useful in the determination of the spectrum of bilinear operators.
By taking a double trace on the forward four-point function, using  \eqref{eq:G-id},
and defining a scalar four-point kernel $K_2$, we write
\be \label{eq:4pt-fwd-ladder-2}
\cF_{ (\mba , \mba ) ; ( \mbm , \mbm )}(x,y,w,z) 
 = N^3 \int_{u,v} \left(\mathbb{I}-K_2\right)^{-1} (x,y,u,v) \left(G(u,w)G(v,z) + G(u,z)G(v,w) \right)\,,
\ee
where
\be \label{eq:K2}
K_2(x,y,w,z) = \int_{u,v} G(x,u) G(y,v) \left( 3\l_t^2 G(u,v)^2 - (\l_p+\l_d) \d(u-v)\right) \d(u-w)\d(v-z) \,.
\ee
Expanding $(\mathbb{I}-K_2)^{-1}$ in a geometric series, one recognizes that \eqref{eq:4pt-fwd-ladder-2} expresses the four-point function as a sum over a mixture of ladder and chain diagrams, built on the building blocks represented in Fig.~\ref{fig:kernel}. For $\l_t=0$ we would have the typical chain diagrams of vector models, while for $\l_p=\l_d=0$ we would have pure ladder diagrams.

The kernel $K_2$ plays an important role in determining the spectrum of operators that appear in the operator product expansion (OPE) of two fundamental fields.
The way this comes about is very similar to the SYK case in $d=1$ \cite{Maldacena:2016hyu}, and it has been discussed in detail in \cite{Liu:2018jhs,Gurau:2019qag}.
The essence is that the forward four-point function can be written in a standard representation-theoretic form \cite{Simmons-Duffin:2017nub}:
\begin{equation} \label{eq:4pt}
\begin{split}
 \cF_{ (\mba , \mba ) ; ( \mbm , \mbm )}(x,y,w,z) 
 = N^3 \sum_{J} 
  \int_{\frac{d}{2}-\imath \infty}^{\frac{d}{2}+\imath\infty} \frac{dh}{2\pi \imath} 
  \;\frac{1}{1-k(h,J)} \; \mu_{\Delta_{\phi}}^d(h,J)
     \cG^{\Delta_{\phi}}_{h,J}(x_i) + (\text{non-norm.})\,,
\end{split}
\end{equation}
with $\cG^{\Delta_{\phi}}_{h,J}(x_i) $ the conformal block, $\mu_{\Delta_{\phi}}^d(h,J)$ the measure, and $k(h,J)$ the eigenvalues of the two particle irreducible four-point kernel $K_2$. The latter can be evaluated explicitly, the corresponding eigenfunctions having the form of three-point functions $\Braket{\phi \phi O_{h,J}}$ of two fundamental fields with an operator of dimension $h$ and spin $J$ \cite{Klebanov:2016xxf,Giombi:2017dtl}.
The non-normalizable contributions are due to operators with dimension $h<d/2$, and they should be treated separately \cite{Simmons-Duffin:2017nub}. The reason for the appearance of $k(h,J)$ should be clear from Eq.\eqref{eq:4pt-fwd-ladder-2}.
Closing the contour to the right, we pick poles at $k(h,J)=1$ (other poles coming from the measure and the conformal block are spurious and they cancel out \cite{Simmons-Duffin:2017nub}), and we recover an operator-product expansion in the $s$-channel ($12\to34$):
\begin{equation}  \label{eq:4pt-OPE}
  \cF_{ (\mba , \mba ) ; ( \mbm , \mbm )}(x,y,w,z) 
 = N^3 \sum_{m,J} c_{m,J}^2  \; \cG^{\Delta_{\phi}}_{h_{m,J},J}(x_i) \,,
\end{equation}
where the dimensions of spin-$J$ operators, $h_{m,J}$, are the poles of $(1-k(h,J))^{-1}$, and the squares of the OPE coefficients $c_{m,J}$ are the residues at the poles \cite{Liu:2018jhs,Gurau:2019qag,Benedetti:2019ikb}.

Therefore, studying the four-point kernel we can obtain the structure of the four-point function, and hence construct renormalized couplings and beta functions, as well as deduce the spectrum of operators that appear in the OPE of two fundamental fields.

%%%%%%%%%%%%%%%%%%%%%%%%
\section{Fixed points in bosonic models}
\label{sec:boson}
%%%%%%%%%%%%%%%%%%%%%%%%

As we discussed, the SD equation \eqref{eq:SDE-p} admits scaling solutions, either asymptotically in the UV or IR (for $\z=1$), or at all energies (for $\z=d/4$). 
However, as any field theory, higher $n$-point functions are typically divergent, hence we need renormalization. Scaling invariance (and likely conformal invariance) for the full theory will be achieved only at a fixed point of the renormalization group. 
This section is devoted to reviewing what we know so far about fixed points for the bosonic quartic model \eqref{eq:int-action-graph}, and its sextic generalizations.

%%%%%%%%%%%%%%%%%%%%%%%%
\subsection{Fixed points {\it \`a la} Wilson-Fisher}
%%%%%%%%%%%%%%%%%%%%%%%%

We first review the status of short-range models, that is, models with free covariance \eqref{eq:free-C} with $\z=1$.
The tensor structure is irrelevant for the standard power counting, which is then the same as for ordinary scalar field theory. Therefore, we know that the critical dimension of quartic models is $d_c=4$, while for sextic models is $d_c=3$, and so on. In this case, fixed points will be found by dimensional continuation below the critical dimension, as in the classical Wilson-Fisher fixed point \cite{Wilson:1971dc}. That is, fixed points are found at small $\epsilon$ in dimension $d=4-\epsilon$ for the quartic model, and so on.

The short-range model with interaction \eqref{eq:int-action-graph} was studied first in $d>0$ (without pillow and double trace terms) by Klebanov and Tarnopolsky in \cite{Klebanov:2016xxf}, where the scaling solution for the two-point function was discussed. As the large-$N$ limit for such model (in $d=0$) was first proved Carrozza and Tanasa in \cite{Carrozza:2015adg}, such model is sometimes called the CTKT model.
Soon after, Giombi et al.\ in \cite{Giombi:2017dtl} developed further such analysis and studied the system of beta functions for the full quartic set of interactions, using the results of \cite{Jack:1990eb} for a theory of scalar fields with a generic quartic potential.
At leading order in $1/N$, and at cubic order in the couplings (i.e.\ at two loops) the beta functions have the following experession:
\begin{align} \label{eq:beta-eps-quartic}
\b_t  &=  -\epsilon g_t +2 g_t^3 \,, \\
\b_p &=  -\epsilon g_p + \left(6 g_t^2 +\f23 g_p^2\right) -2 g_t^2 g_p\,, \\
\b_d &=  -\epsilon g_d + \left(\f43 g_p^2 + 4 g_p g_d +2 g_d^2\right) -2 g_t^2 \left(4 g_p +5g_d\right)\,,
\end{align}
where we denote by $g_i$, with $i=t,p,d$, the renormalized couplings, rescaled by $(4\pi)^2$.

The beta function of $g_t$ depends only on $g_t$ itself, and in fact it is entirely due to the wave function renormalization. That is, the tetrahedron coupling receives no vertex corrections at leading order in $1/N$; this facts holds at all loop orders, as can be proved by analyzing the structure of the four-point function \eqref{eq:4pt-fwd-ladder} \cite{Benedetti:2019eyl}.

Notice also that the double-trace coupling only enters in its own beta function. This can in part be understood as follows: setting $g_t=g_p=0$ we obtain a vector model in disguise, with $O(N^3)$ invariance. Such enhanced symmetry prevents the generation of the pillow and tetrahedron interactions if they are initially absent, hence 
there cannot be pure $g_d$ terms in their beta functions. That there are also no mixed terms involving $g_d$, at any loop order, can again be proved by analyzing the structure of the four-point function \eqref{eq:4pt-fwd-ladder} \cite{Benedetti:2019eyl}.

The beta functions \eqref{eq:beta-eps-quartic} admit several fixed points: the trivial one, the $O(N^3)$ Wilson-Fisher fixed point, two Wilson-Fisher-like with $g_t=0$ but non-zero pillow, and lastly eight non-trivial fixed points with $g_t\neq 0$, i.e.\
\be
g_t^* = \pm \sqrt{\eps/2} \,, \;\;\; g_p^* = \pm3\im \sqrt{\eps/2}\,, \;\;\; g_d^*\mp \im (3\pm\sqrt{3}) \sqrt{\eps/2} \,,
\ee
where for the global sign of $g_d^*$ the choice is synchronized with that for $g_p^*$.
It is the latter fixed points that we call \emph{melonic}, as only for non-zero tetrahedron coupling we have melonic diagrams at large-$N$.
The critical exponents (i.e.\ the eigenvalues of the stability matrix $\cB_{ab}=\p_{g_a}\b_b|_{g=g^*}$) of the melonic fixed points are
\be
\{\, 2\eps\,, \; \pm 2\im \sqrt{2\eps} \,, \; \pm 2\im \sqrt{6\eps}\, \} \,,
\ee
with corresponding right-eigenvectors (at leading order in $\eps$)
\be
\{\, (1,0,0)\,,\; (\pm \im 3,1,0)\,,\; (\pm \im \sqrt{3},1,1)  \, \} \,.
\ee
The first exponent is real positive, and with eigenvector $(1,0,0)$, meaning that tetrahedron perturbations are irrelevant, i.e.\ IR attractive. 
However, the other two exponents are purely imaginary, meaning that in the (complex) plane defined by the other two eigendirections the trajectories circle around the fixed point without ever reaching it.

Moreover, the complex critical exponents mean that the fixed-point theory has operators of complex dimension, hence it is non-unitary and likely unstable. 
The instability can  be understood from an AdS/CFT point of view, as  the critical exponent $\pm 2\im \sqrt{6\eps}$ gives a conformal dimension $\Delta_{\phi^2}=\f{d}{2}\pm \im \sqrt{6\eps}$ for the operator $\phi_{abc}\phi_{abc}$, implying a violation of the Breitenlohner-Freedman stability bound in AdS${}_{d+1}$ \cite{Breitenlohner:1982jf}.
In this respect, it was conjectured in \cite{Kim:2019upg} that ``if the assumption of conformal invariance in a large-$N$ theory leads to a single-trace operator with a complex scaling dimension of the form $\f{d}{2} +\im f$, then in the true low-temperature phase this operator acquires a vacuum expectation value'', thus spontaneously breaking conformal invariance.

The appearance of an instability is not unexpected, as the tetrahedron invariant is not positive definite. However, the unstable potential does not prevent the theory to be defined perturbatively, and in principle in the large-$N$ limit the instability of the potential might be invisible (in $d=0$ this is very familiar, as critical points of matrix models are usually found at the ``wrong'' sign of the coupling \cite{Brezin:1977sv}). The above result shows that this is not the case for such type of melonic fixed point.

Notice that the melonic fixed points above are real, with real exponents, for $\eps<0$, i.e.\ above the critical dimension. However, in such case the tetrahedron is always a relevant perturbation, that is, the fixed point is UV attractive. Moreover, complex dimensions reappear for $d>4.155$ \cite{Giombi:2017dtl}.

\paragraph{Sextic models.} One can generalize the two-point function and beta functions analysis to sextic potentials. Sextic models have first been considered formally in \cite{Giombi:2017dtl}, while beta functions have been studied in \cite{Giombi:2018qgp,Benedetti:2019rja}.
The model considered in \cite{Giombi:2018qgp} has $O(N)^3$ invariance, but a different scaling than \eqref{eq:opt-scal} has been chosen. The sextic invariant leading to melonic diagrams in that case has 3-colored graph in the form of a prism, hence we will call this model the \emph{prismatic model}.
The model studied in \cite{Benedetti:2019rja} has instead $U(N)^3$ invariance, and with scaling parameters \eqref{eq:opt-scal}. The invariant leading to melonic diagrams in that case is the complete bipartite graph $K_{3,3}$, resembling a wheel, hence we call this model the \emph{wheel model}.
More explicitly, the latter has the following interacting action:
\be
\begin{split} \label{eq:6int-action-graph}
S_{\rm int}[\phi,\phib] = & \int d^d x  
\left( \f{\l_1}{6 N^{3}} \vcenter{\hbox{\includegraphics[width=1.6cm]{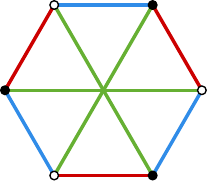}}}
+ \f{\l_2}{6 N^{4}} \vcenter{\hbox{\includegraphics[width=1.6cm]{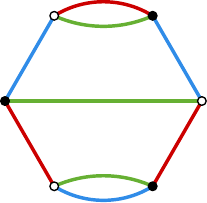}}}
+ \f{\l_3}{6 N^{4}} \vcenter{\hbox{\includegraphics[width=1.6cm]{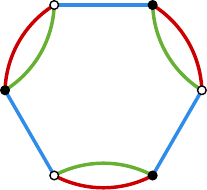}}}
 \right. \\
 & \left.
 + \f{\l_4}{6 N^{5}} \vcenter{\hbox{\includegraphics[width=.9cm]{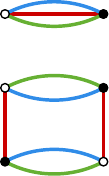}}}
  + \f{\l_5}{6 N^{6}} \vcenter{\hbox{\includegraphics[width=.9cm]{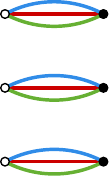}}}
  \right)\,,
\end{split}
\ee
where a (normalized) sum over color permutations should be understood, whenever it is non-trivial.

For both the prismatic and the wheel model, real melonic fixed points have been found, and both the critical exponents and the spectrum of bilinear operators are real for small $\epsilon$. However, complex dimensions reappear in the two models for $\eps\simeq 0.19$ and $\eps\simeq 0.02$, respectively.

Ref.~\cite{Benedetti:2019rja}  also considered an $O(N)^5$ model, with a restricted (``melo-complete'') family of invariants, which is however closed under renormalization at large $N$. Again melonic diagrams dominate at large-$N$, due to the invariant corresponding to the complete graph $K_6$. However, in this case no non-trivial fixed point was found.
Notice that this rank-5 model was the one considered formally  in \cite{Giombi:2017dtl}, and for which a non-trivial spectrum of bilinear operators was found. The result of \cite{Benedetti:2019rja} thus shows the importance of considering the full set of beta functions.

%%%%%%%%%%%%%%%%%%%%%%%%
\subsection{Lines of fixed points in long-range models}
%%%%%%%%%%%%%%%%%%%%%%%%

Partially inspired by \cite{Gross:2017vhb}, a long-range version of the CTKT model was introduce in \cite{Benedetti:2019eyl}.
The model has the same potential \eqref{eq:int-action-graph}, but in the free part of the action one chooses $\z=d/4$, a value that in the long-range Ising model ($N=1$) corresponds to the transition point between mean-field and non-trivial long-range behavior.
For $d<4$ the kinetic term is nonlocal, and thus one finds that there is no wave function renormalization. As we explained above, at large-$N$ the wave function renormalization is the only possible reason for a renormalization group flow of the tetrahedron coupling, as this receives no radiative correction at leading order; therefore, in the long-range case one finds that the tetrahedron coupling is exactly marginal at large-$N$.
As a consequence, we can consider $d$ smaller and not close to 4, for example $d=3$ or $d=2$, while still having the luxury of a small parameter, which this time is the marginal coupling $g_t$. The combination of large-$N$ and small $g_t$ gives full control on the fixed points of the theory in any $d<4$.

Surprisingly, it turns out that the fixed points and their critical exponents, as well as the scaling dimensions and OPE coefficients appearing in \eqref{eq:4pt-OPE},  are all real at small $g_t$, if one chooses the latter to be purely imaginary \cite{Benedetti:2019eyl,Benedetti:2019ikb,Benedetti:2020yvb}.
Notice that since the tetrahedron invariant is not positive definite, the choice of imaginary coupling is almost mandatory from the  point of view of a non-perturbative finite-$N$ functional integral, and it is also reminiscent of the Lee-Yang model with an $\im \lambda \phi^3$ interaction \cite{Fisher:1978pf,Cardy:1985yy}.

Defining the new couplings $\lambda_1 = \lambda_p/3$, $\lambda_2 = \lambda_p + \lambda_d$, and  $\l=-\im\l_t$, the beta functions of the respective renormalized couplings $g_1$ and $g_2$ are
  \begin{align}
 \beta_{\gt_1} & %=  k\frac{ \partial \gt_1}{ \partial k}\Big{|}_{\lambda,\lambda_1}     
   =  \beta_0^{\gt} - 2 \beta_1^{\gt} \,  \gt_1 +
   \beta_2^{\gt} \, \gt_1^2 \,, 
    \crcr
  \beta_{\gt_2} & %=  k\frac{ \partial g_2}{ \partial k}\Big{|}_{\lambda,\lambda_2}     
   =  \beta_0^{ \sqrt{3} \gt }- 2 \beta_1^{\sqrt{3} \gt } \,  \gt_2 +
   \beta_2^{ \sqrt{3} \gt }\,  \gt_2^2 \,,
  \end{align}
where $   \beta_0^{ g}, \beta_1^{ g } $  and   $\beta_2^{g}$ are power series in $g^2$, with $g=Z^{-2}\l$ and $Z$ solution of \eqref{eq:wave}, and we have rescaled the couplings to $\tilde{g}_i = g_i (4\pi)^{- d/2} \Gamma(\zeta)^{-2} $. 
At first order the coefficients are
\begin{align}
 \beta_0^{\gt} = 
 - \left( 2\frac{\Gamma(\frac{d}{4})^2}{ \Gamma(\frac{d}{2})}\right) \gt^2+  \mathcal{O}(\gt^4) \,,\qquad
 \beta_1^{ \gt} =  \mathcal{O}(\gt^2) \,, \qquad
  \beta_2^{ \gt} = \left( 2\frac{\Gamma(\frac{d}{4})^2}{ \Gamma(\frac{d}{2})}\right) + \mathcal{O}(\gt^2) \,.
\end{align}

The beta function $\beta_{\gt_1}$ admits two fixed points:
   \begin{align} \label{eq:FP}
 \gt_{1\pm} =  \frac{
    \beta_1^{\gt} 
    \pm \sqrt{ (\beta_1^{\gt})^2 -\beta_0^{\gt}\beta_2^{\gt} }
    }{\beta_2^{\gt}}  = \pm\sqrt{\gt^2} +  \mathcal{O}(\gt^2) \; ,
   \end{align} 
 and the beta function $\beta_{\gt_2}$ admits two fixed points of the same form, with $\gt\to\sqrt{3}\gt$.
 The corresponding critical exponents are
 \begin{align}  \beta'_{\gt_1}( \gt_{1\pm}) &= \pm 2\sqrt{ (\beta_1^{\gt})^2 -\beta_0^{\gt}\beta_2^{\gt} }
     = \pm \sqrt{\gt^2} \left( 4\frac{\Gamma(\frac{d}{4})^2}{ \Gamma(\frac{d}{2})}\right) +  \mathcal{O}(\gt^3)\,, \\
     \label{eq:exp}
  \beta'_{\gt_2}( \gt_{2\pm}) &= \pm 2\sqrt{ (\beta_1^{\sqrt{3}\gt})^2 -\beta_0^{\sqrt{3}\gt}\beta_2^{\sqrt{3}\gt} }
     = \pm \sqrt{3 \gt^2} \left( 4\frac{\Gamma(\frac{d}{4})^2}{ \Gamma(\frac{d}{2})}\right) +  \mathcal{O}(\gt^3)\,,
 \end{align}
 which are real for $\gt\in\mathbb{R}$, i.e.\ $\l_t$ purely imaginary.
 Hence, the model has four fixed points in total, each of them actually defining a line parameterized by $\gt$ in the $\{\gt_1,\gt_2\}$ plane.
One of them, namely $\{\gt_{1+},\gt_{2+}\}$, is IR attractive in both directions.
 For $\gt\to 0$, they all merge into the trivial fixed point: for $g=0$, non-trivial fixed points can only be obtained by moving away from marginality (i.e.\ by taking $4\zeta-d=\epsilon>0$).

As argued in \cite{Benedetti:2020yvb}, following the footsteps of \cite{Paulos:2015jfa}, the fixed-point theory is not only scale invariant, as any fixed-point theory, but also conformally invariant.

The spectrum of dimensions of bilinear operators of arbitrary spin, as well as their OPE coefficients, and some OPE coefficients of quartic operators, have been computed in 
\cite{Benedetti:2019eyl,Benedetti:2019ikb,Benedetti:2020yvb}, by the method outlined in Sec.~\ref{sec:4pt-K}.
The eigenvalues of the two particle irreducible four point kernel \cite{Gurau:2019qag} are
\begin{equation}
k(h,J)=3g^2\Gamma(d/4)^4
 \frac{\Gamma(-\frac{d}{4}+\frac{h+J}{2})\Gamma(\frac{d}{4}-\frac{h-J}{2})}{\Gamma(\frac{3d}{4}-\frac{h-J}{2})\Gamma(\frac{d}{4}+\frac{h+J}{2})} \,,
\label{eq:k(h,J)}
\end{equation}
from which one finds \cite{Benedetti:2019ikb} two types of solutions of the equation $k(h,J)=1$ at small renormalized tetrahedral coupling $g$. The first type,
\begin{equation} \label{eq:h_pm}
h_{\pm} = \frac{d}{2} \pm 2 \frac{  \Gamma(d/4)^2  }{ \Gamma(d/2) } \sqrt{ 3g^2  } + {\cal O}(g^3) \, ,
\end{equation}
only exists for the scalar (spin $J=0$) case. It is real (at all orders in $g$) only for real $g$, i.e.\ for purely imaginary tetrahedral coupling. 
Moreover we recognize that $h_{\pm} =\f{d}{2}+\f12 \beta'_{\gt_2}( \gt_{2\pm})$, as the expected dimension of the composite operator $\phi_{abc} \phi_{abc}$ in the large-$N$ limit (that is, half the dimension of $(\phi_{abc} \phi_{abc})^2$).
The second type of solution,
\begin{equation} \label{eq:h_mJ}
 h_{m,J} = \frac{d}{2} + J + 2m -  \frac{ \Gamma(d/4)^4 
     \Gamma(m + J)  \Gamma(m+1 - \frac{d}{2}) \sin\left(\frac{\pi d}{2} \right)  }{ \Gamma(\frac{d}{2} + J + m ) \Gamma(m+1) \; \pi } 6g^2  + 
     {\cal O}(g^4) \, ,
\end{equation}
with $m,J \in \mathbb{N}_0$, but not simultaneously zero, exists for both scalar $J=0$ and spin $J>0$. It is real (at all orders in $g$) for both real and purely imaginary tetrahedral coupling. 
In the free limit $g=0$, we recover the classical dimensions $\frac{d}{2}+J+2m$ of the primary bilinear operators with arbitrary spin $J$, schematically of the form
\be
O_{h,J} \sim  \phi_{abc} \partial_{\mu_1}\dots \partial_{\mu_J} (\partial^2)^{m}\phi_{abc}  \,.
\ee

The solutions \eqref{eq:h_pm} and \eqref{eq:h_mJ} are all real for real $g$, and above unitarity bounds (e.g.\ \cite{Poland:2018epd}), and the associated OPE coefficients are also real \cite{Benedetti:2019ikb}.
This raises the possibility that, despite the imaginary coupling, for which we expect the full model to be non-unitary, at large-$N$ it could actually be unitary. A possible mechanism for such outcome could be that the non-unitarity would for example manifest itself in the form of complex operator dimensions, but with imaginary part suppressed in the large-$N$ limit, $h\simeq \a_0 + \im \a_1/N +{\rm O}(1/N^2)$.

A similar scenario is found also in the long-range version of the sextic wheel model \eqref{eq:6int-action-graph} \cite{Benedetti:2019rja}.
Choosing $\z=d/3$ and $d<3$, one finds a line of IR attractive fixed points parametrized by $g_1$. However, in this case critical exponents and scaling dimensions of bilinear operators are real for real $g_1$, despite the fact that the wheel invariant is also not positive definite.

%%%%%%%%%%%%%%%%%%%%%%%%
\section{Fermionic models}
\label{sec:fermion}
%%%%%%%%%%%%%%%%%%%%%%%%

Fermionic models with a melonic limit have been the original driving force for exploring tensor models in $d>0$. 
The SYK model \cite{Sachdev:1992fk,Kitaev} is a model of Majorana fermions in $d=1$, which, thanks to the melonic dominance, admits a conformal IR limit and has an out-of-time-order correlator that saturates the chaos bound \cite{Polchinski:2016xgd,Maldacena:2016hyu}, and as a consequence it attracted much attention.
However, the SYK model is not a proper quantum field theory (or rather quantum mechanical model), as its coupling is randomly distributed and one needs to perform a quenched average on free energy and correlators. Moreover, the model has no global symmetry at the fundamental level, it is a model of $N$ fermions with a completely generic (randomly distributed) $q$-body interaction. A global $O(N)$ symmetry only arises after the quenched/annealed average on the coupling, hence it cannot be gauged, and therefore its restriction to the singlet sector (in view of an AdS/CFT correspondence) remains rather artificial.
For these reasons, the observation \cite{Witten:2016iux,Klebanov:2016xxf} that the same conformal limit and correlator could be obtained in a proper field theory model with a global invariance has attracted some attention on tensor models.
We refer to \cite{Delporte:2018iyf,Klebanov:2018fzb} for a review of SYK-like tensor models ($d=1$).
This section is devoted to briefly reviewing few facts we have learned so far about fermionic tensor field theories in $d>1$ dimensions.

%%%%%%%%%%%%%%%%%%%%%%%%
\subsection{Tensorial Gross-Neveu models}
%%%%%%%%%%%%%%%%%%%%%%%%

%We first consider the $d=2$ model of 
Fermionic tensor field theory with quartic interactions have been studied in \cite{Prakash:2017hwq,Benedetti:2017fmp,Benedetti:2018ghn}.
Unlike in $d=1$, fermions in higher dimensions have a spinor structure that introduces some new features in the construction of invariants. 
This aspect has been investigated in some detail in \cite{Benedetti:2017fmp},  for $d=2$, which  is the critical dimension for fermions with standard propagator and quartic interactions. We will first review the construction for the $O(N)^3$-invariant model in $d=2$  \cite{Benedetti:2017fmp}, and then comment on $d\neq 2$ and other symmetries.

The free part of the action is standard, as there is only one quadratic tensor invariant, like for vectors:
\be \label{eq:free-M}
S_{\rm free} =  \f12 \int_x\; \psib_{abc}(x) \slashed{\p} \psi_{abc}(x)  \,.
\ee
The field $\psi_{abc}(x)$ is a real Majorana spinor, but it could be taken to be a Dirac spinor in the case of $U(N)^3$ symmetry, or of a mixture of the two cases.
The inclusion of a quartic potential leads to a tensorial generalization of the Gross-Neveu model \cite{Gross:1974jv}.

\begin{figure}[ht]
 \centering
        \begin{minipage}{0.25\textwidth}
            \centering 
             \includegraphics[scale = 0.65]{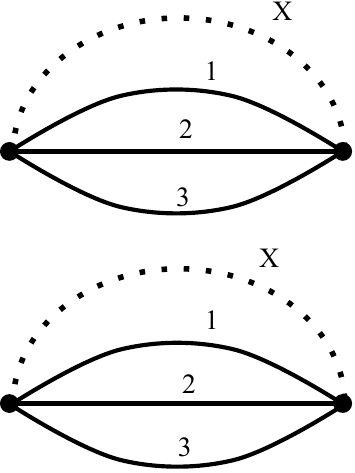}
        \end{minipage}
        \hspace{0.01\textwidth}
        \begin{minipage}{0.25\textwidth}
            \centering
            \includegraphics[scale = 0.65]{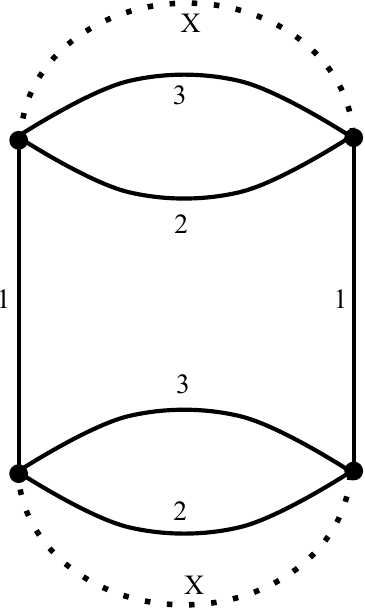}
        \end{minipage}
        \hspace{0.01\textwidth}
        %\\
        %\vspace{.4cm}
        \begin{minipage}{0.25\textwidth}
            \centering
            \includegraphics[scale = 0.65]{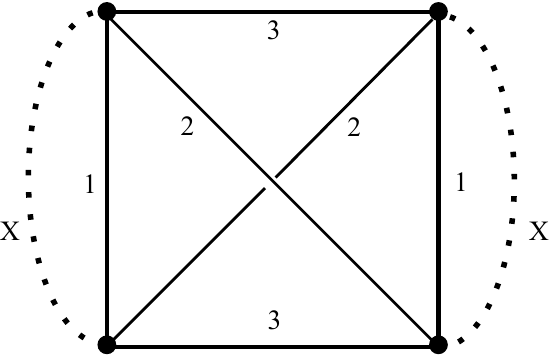}
        \end{minipage}
  \caption{\small{Graphical representation of the interaction vertices \eqref{eq:vertex0} (left), \eqref{eq:vertex1} (center), and \eqref{eq:vertex2} (right). 
  In place of colors, here we used labels $i\in\{1,2,3\}$, representing the contraction of the $i$-th  tensorial indices of the two fields at its end-vertices. A dotted line with label $X$ represents instead the contraction of the spinorial indices of the fields at  its end-vertices, with the insertion of a matrix $\G^X$. 
  }}
\label{fig:vertices-fermi}
\end{figure}

The generalization of the quartic potential \eqref{eq:int-action-graph} to the fermionic case requires some care, as in between a $\psib$ and a $\psi$ field we can insert gamma matrices.
We define the matrices $\Gamma^X$, with $X=S,V,P$,  as
\be
\Gamma^S=1, \;\;\; \Gamma^V=\g^\m,
\;\;\; \Gamma^P=\g_5 \,. 
\ee
The letters $S$, $V$, and $P$ stand for scalar, vector, and pseudoscalar, respectively, in reference to the transformation properties of the associated bilinears under the rotation group.
The double-trace and pillow invariants now write
\begin{align}
 \label{eq:vertex0}
 I_d^X = &  (\psib_{a_1 a_2 a_3} \Gamma^X \psi_{a_1 a_2 a_3})( \psib_{b_1 b_2 b_3} \Gamma^X \psi_{b_1 b_2 b_3} )\,,\\
 \label{eq:vertex1}
\begin{split}
I_p^{X,1} = 
& (\psib_{a_1a_2 a_3} \Gamma^X \psi_{b_1 a_2 a_3} )(\psib_{b_1 b_2 b_3} \Gamma^X \psi_{a_1 b_2 b_3} )\; ,
\end{split}
\end{align}
where terms inside a parenthesis have all their spinorial indices contracted, and the index 1 in the pillow invariant stands for the color of the vertical line in its graph representation, see Fig.~\ref{fig:vertices-fermi}.
Demanding symmetry under color permutations requires repackaging the pillow interactions as
%
%\be
$I_p^X = \sum_{\ell=1}^3 I_p^{X,\ell}$, as in \eqref{eq:int-action-graph}.
%\ee
%
However, in the fermionic case also the tetrahedron invariant depends on the choice of coloring; it turns out that there is no tetrahedron interaction invariant under the full permutation group, but we can instead introduce an invariant under the subgroup of even permutations:
\be \label{A3_inv}
I_t \equiv - I_{t}^{V, \ell} + 2 I_{t}^{P, \ell} \,,
\ee
where
\begin{align} \label{eq:vertex2}
I_{t}^{X,1} &= \big(\psib_{a_1 a_2 a_3} \Gamma^X \psi_{a_1 b_2 b_3}\big)\,\big( \psib_{b_1 a_2 b_3} \Gamma^X \psi_{b_1 b_2 a_3} \big)\,,\crcr
I_{t}^{X,2} &= \big(\psib_{a_1 a_2 a_3} \Gamma^X \psi_{b_1 a_2 b_3}\big)\,\big( \psib_{b_1 b_2 a_3} \Gamma^X \psi_{a_1 b_2 b_3} \big)\,, \\
I_{t}^{X,3} &= \big(\psib_{a_1 a_2 a_3} \Gamma^X \psi_{b_1 b_2 a_3}\big)\,\big( \psib_{a_1 b_2 b_3} \Gamma^X \psi_{b_1 a_2 b_3} \big)\,. \nonumber
\end{align} 
Despite appearances, it can be proved that $I_t$ is independent of $\ell$. Moreover, by use of Fierz identities, one can show that we are free to parametrize a general action in terms of $I_{t}^{X,\ell}$, with an $\ell$ of our choice.
%
%%%%%%%%%%
%%%%%%%%%%

Since $I_d^P=I_d^V=0$ for Majorana fermions, the most general action for quartic interactions invariant under $O(N)^3$ and even color permutations is
\be \label{eq:general-int-M}
S_{\rm int} = - \f14 \int_x \left(\f{\l_d}{N^3}  I_d^S+ \sum_{X=S,V,P}  \f{\l_p^X}{N^2} I_p^X +\f{\l_t}{N^{3/2}}  I_t  \right)\,.
\ee
According to the analysis of  \cite{Benedetti:2017fmp}, there is no non-trivial CFT for such model in $d=2$, which displays a dynamical mass generation, like the usual (vectorial) Gross-Neveu model \cite{Gross:1974jv}, with the non-perturbatively generated mass satisfying
\be
\lim_{\l_t\to 0} m =  \L \exp\left( - \f{\pi}{b^2 (\l_d^S+3\l_p^S)}  \right) \,,
\ee
where $\L$ is the UV cutoff, and $b$ a constant.

It is also interesting to consider a model with no color-permutation symmetry. The most generic action with only $I_2$ interactions reads
\be\label{S2_general}
S_{\mathrm{int},2 }= - \sum_{X=S,V,P} \lambda_t^{X,1} \int_x I_2^{X,1}\,,
\ee
where we chose the $\ell=1$ representation for definiteness. 
In this case, by a sort of bootstrap procedure, it was found in \cite{Benedetti:2017fmp} that requiring no radiative generation of the pillow and double-trace interactions, and requiring no mass generation, leads to one of the following two sets of conditions on the tetrahedron couplings:
\be\label{eq:I2_cond1}
(\lambda_t^{S,1})^2 = (\lambda_t^{P,1})^2 = \lambda^2 \neq 0 \qquad \mathrm{and} \qquad \lambda_t^{V,1} = 0\,,  
\ee
or 
\be\label{eq:I2_cond2}
(\lambda_t^{S,1})^2 = (\lambda_t^{P,1})^2 = 0 \qquad \mathrm{and} \qquad \lambda_t^{V,1} = \lambda \neq 0 \,.
\ee
Interestingly, such conditions are satisfied by models invariant under the continuous chiral symmetry.
However, this is no guarantee of a scale invariant theory, as there is still a possible running of the coupling $\l$, which in the large-$N$ limit is again driven only by the wave function renormalization.
By explicit computation \cite{Benedetti:2017fmp}, the coupling $\l$ in equation \eqref{eq:I2_cond1} has, at leading order in $1/N$ and $\l$, the following beta function:
\be \label{eq:beta2}
\b_\l =  \f{3}{\pi^2} \l^3 \,.
\ee
As a consequence, unlike the usual Gross-Neveu coupling, the coupling $\l$ is not asymptotically free, it is IR free for both signs of the coupling. Such a result derives from the fact that at leading order in $1/N$ the beta function starts at two-loop order (from the melonic correction to the two-point function). The sign in  \eqref{eq:beta2} is thus compatible with results from the usual Gross-Neveu model, where the leading terms in $1/N$ of the two loop contributions to the beta functions have the same positive sign as our \eqref{eq:beta2} (see for example \cite{Bondi:1989nq}). The key difference is that for the vectorial Gross-Neveu model the two-loop part of the beta function is subleading in $1/N$ with respect to the one-loop part, while here the situation is reversed.

Interestingly, in $d=2-\eps$ dimensions the beta function \eqref{eq:beta2} becomes
\be
\b_2 = -\epsilon \l+ \f{3}{\pi^2} \l^3 \,,
\ee
thus showing for $\eps>0$ a non-trivial IR fixed point of order $\sqrt{\epsilon}$, both at positive and negative coupling.
On the contrary, no real UV fixed point is found in $d=2+\epsilon$ dimensions. The situation is thus reversed with respect to the GN model.
Such result provides support for the conjecture in  \cite{Prakash:2017hwq}  that the theory in $d=2-\eps$ dimensions flows to a weakly interacting IR fixed point for small $\epsilon$.
It is also natural to think that such IR fixed point in $d=2-\eps$ would continue to $d=1$, where one would recover the SYK-like conformal limit.

\paragraph{New patterns of symmetry breaking.}
In the case of complex fermions with symmetry $U(N)^3$, the tetrahedron interaction is not allowed.
Therefore in the quartic model we are left with only pillow and double-trace, and 
the large-$N$ limit is dominated by cactus diagrams, as the vector model.
Nevertheless, it is worth to mention a new feature of such model, namely the appearance in $d=3$  of a new phase with spontaneous symmetry breaking of one of the $U(N)$ components of the invariance group.
Spontaneous symmetry breaking can be studied in this case by the intermediate field method (or Hubbard-Stratonovich transformation), similarly to the standard Gross-Neveu case, except that the pillow interaction requires a matrix intermediate field for each pillow invariant (i.e.\ one for each coloring of the pillow graph, or one for each $U(N)$ component of the symmetry group). % {\color{red} [cite d=0 case]}.
In the large-$N$ limit, the effective potential for the intermediate field can be computed, and it was observed first in \cite{Benedetti:2017fmp} (at $d=2$) that one of the intermediate matrix fields could acquire a vacuum expectation value proportional to a non-trivial traceless matrix, thus breaking one of the $U(N)$ components of the symmetry group.
However, the Coleman-Mermin-Wagner theorem forbids spontaneous symmetry breaking of continuous symmetries in $d=2$, therefore the apparent breaking found in  \cite{Benedetti:2017fmp} is a spurious effect of the large-$N$ limit, similar to the apparent breaking of continuous chiral symmetry in the Gross-Neveu model \cite{Gross:1974jv,Witten:1978qu}.
On the other hand, the generalization of the model to $d=3$, where there is no obstruction to spontaneous symmetry breaking of continuous symmetries, confirmed such pattern of symmetry breaking \cite{Benedetti:2018ghn}. The $d=3$ model is of course perturbatively non-renormalizable, however, as the original Gross-Neveu model, it makes sense in the large-$N$ limit; moreover, it admits three non-trivial fixed points, one UV attractive in both directions of the $\{\l_p,\l_d\}$ plane, and two saddle-type fixed points. However, all such fixed points are similar to the usual vectorial Gross-Neveu fixed points and thus do not fit our definition of melonic CFTs.

%%%%%%%%%%%%%%%%%%%%%%%%
\section{Conclusions and outlook}
\label{sec:concl}
%%%%%%%%%%%%%%%%%%%%%%%%

We hope to have managed to convey some essential information on the status of tensor field theories in the melonic limit.
Our presentation has been limited in scope, omitting for example any discussion of the $d=1$ case or of supersymmetric models (see \cite{Peng:2016mxj,Chang:2018sve,Popov:2019nja} for the latter).
We have tried instead to highlight some recurrent features, such as:
\begin{itemize}
\item the general structure of the invariants, their graphical representation, and the counting of factors of $N$ in the Feynman diagrams;

\item the role of the large-$N$ melonic dominance in the simplification of the Schwinger-Dyson equations for the two-point function, which can be fully solved in the critical long-range models, and of the four-point kernel, whose spectrum can be solved numerically;

\item the two main mechanisms by which fixed points can appear, namely: the usual Wilson-Fisher dimensional continuation below the upper critical dimension in the short-range models, or the exact marginality of the maximally-single-trace interaction in the long-range models. 

\item the fact that complex scaling dimensions often show up, but for some models small ranges of parameters (the dimension or the exactly marginal coupling) have been found for which all the scaling dimensions (and OPE coefficients) computed so far are real and above unitarity bounds.

\end{itemize}

The long-range models  \cite{Benedetti:2019eyl,Benedetti:2019ikb,Benedetti:2020yvb} are of particular interest, as they allow us to define real melonic CFTs in some integer dimensions (such as $d=2$ or $d=3$) for a small-enough exactly marginal coupling. However, we expect that the exact marginality will be lost at some higher order in the $1/N$ expansion, and thus it is a pressing question to understand whether the $1/N$ corrections will select at least one fixed-point value within the range in which the model CFT is real.
It would also be interesting to study fermionic long-range models \cite{Gawedzki:1985ed,Gawedzki:1985jn} with tensor invariance; these might provide a higher-dimensional, and tensorial, generalization of the conformal SYK model of Ref.~\cite{Gross:2017vhb}.

It is also worth to mention some similarities to other models, as well as other generalizations.
As recently noticed in \cite{Benedetti:2020yvb}, the quartic long-range model has several similarities with the long-range version \cite{Kazakov:2018qbr} of fishnet theory \cite{Gurdogan:2015csr}. This is due in particular to the structure of the four-point function, renormalizing the double-trace (and pillow, in the tensor case) interactions: the ladder structure indeed can originate both from opening two lines in melonic vacuum graphs, or from considering a fishnet graph with only four external legs. The double-trace (and pillow) couplings thus present in both models similar lines of fixed points, parametrized by the exactly marginal coupling. Moreover, in both models the latter is associated with a complex operator: $i$ times the tetrahedron invariant in the tensor case, and a ``chiral'' vertex in the fishnet case. Therefore, both models are expected to be non-unitary. However, while the non-unitarity has showed up in the fishnet theory even at leading order (in the form of a logarithmic CFT type of correlators), the large-$N$ results in the long-range tensor models (in the appropriate range) are so far compatible with a unitary CFT.
 
An interesting generalization of tensor models which we didn't discuss is that introduced by Ferrari in \cite{Ferrari:2017ryl}, and further studied in \cite{Ferrari:2017jgw,Azeyanagi:2017drg,Azeyanagi:2017mre,Benedetti:2020iyz,Carrozza:2020eaz}, in which only two indices of a tensor field take values from 1 to $N$, while one index (or more, for higher rank than three) takes values from 1 to $D$. This allows to view the melonic limit as a new limit (large $N$ and $D$) of multi-matrix models, with possibly an easier link to the AdS/CFT correspondence. In this respect it is also interesting to notice that a similar generalization of the sextic wheel model of Ref.~\cite{Benedetti:2019rja}, with $D=4$, essentially reduces to the bosonic potential of the ABJM model \cite{Aharony:2008ug}, an observation that might be worth of some exploration.

Many interesting questions concerning tensor field theories are still open, such as the determination of conformal data for higher-order invariants, the fate of the fixed points at subleading order in $1/N$, and a possible AdS/CFT interpretation of the models.

%%%%%%%%%%%%%%%%%%%%%%%%
\acknowledgments
I wish to thank the organizers of CORFU2019 for the very enjoyable workshop. I also wish to thank Sylvain Carrozza, Nicolas Delporte, Razvan Gurau, Sabine Harribey, Alessandro Sfondrini, Ritam Sinha, and Kenta Suzuki for collaboration on the joint works presented here.
I acknowledge support by the European Research Council (ERC) under the European Union's Horizon 2020 research and innovation program (grant agreement No818066).

\newpage
%%%%%%%%%%%%%%%%%%%%%%%%

\providecommand{\href}[2]{#2}\begingroup\raggedright\endgroup


\begin{thebibliography}{100}

\bibitem{Wilson:1972cf}
K.~G. Wilson, \emph{Quantum field theory models in less than four-dimensions},
  \href{https://doi.org/10.1103/PhysRevD.7.2911}{\emph{Phys.Rev.D} {\bfseries
  7} (1973) 2911}.

\bibitem{Coleman:1974jh}
S.~R. Coleman, R.~Jackiw and H.~Politzer, \emph{Spontaneous symmetry breaking
  in the {O(N)} model for large {N}},
  \href{https://doi.org/10.1103/PhysRevD.10.2491}{\emph{Phys.\ Rev.\ D}
  {\bfseries 10} (1974) 2491}.

\bibitem{Gross:1974jv}
D.~J. Gross and A.~Neveu, \emph{Dynamical symmetry breaking in asymptotically
  free field theories},
  \href{https://doi.org/10.1103/PhysRevD.10.3235}{\emph{Phys. Rev.} {\bfseries
  D10} (1974) 3235}.

\bibitem{Coleman:1985rnk}
S.~Coleman, \emph{{Aspects of symmetry: selected Erice lectures}}. Cambridge
  University Press, Cambridge, U.K., 1985,
  \href{https://doi.org/10.1017/CBO9780511565045}{10.1017/CBO9780511565045}.

\bibitem{'tHooft:1973jz}
G.~'t~Hooft, \emph{{A planar diagram theory for strong interactions}},
  \href{https://doi.org/10.1016/0550-3213(74)90154-0}{\emph{Nucl. Phys.}
  {\bfseries B72} (1974) 461}.

\bibitem{Brezin:1977sv}
E.~Brezin, C.~Itzykson, G.~Parisi and J.~B. Zuber, \emph{{Planar diagrams}},
  \href{https://doi.org/10.1007/BF01614153}{\emph{Commun. Math. Phys.}
  {\bfseries 59} (1978) 35}.

\bibitem{Brezin:1994eb}
E.~Brezin and S.~Wadia, eds., \emph{{The Large N expansion in quantum field
  theory and statistical physics: From spin systems to two-dimensional
  gravity}}. World Scientific, Singapore, 1994.

\bibitem{BenGeloun:2011rc}
J.~Ben~Geloun and V.~Rivasseau, \emph{{A renormalizable 4-Dimensional tensor
  field theory}},
  \href{https://doi.org/10.1007/s00220-012-1549-1}{\emph{Commun. Math. Phys.}
  {\bfseries 318} (2013) 69} [\href{https://arxiv.org/abs/1111.4997}{{\ttfamily
  1111.4997}}].

\bibitem{BenGeloun:2012pu}
J.~Ben~Geloun and D.~O. Samary, \emph{{3D tensor field theory: renormalization
  and one-loop {$\beta$}-functions}},
  \href{https://doi.org/10.1007/s00023-012-0225-5}{\emph{Ann. H. Poincar\'e}
  {\bfseries 14} (2013) 1599}
  [\href{https://arxiv.org/abs/1201.0176}{{\ttfamily 1201.0176}}].

\bibitem{BenGeloun:2012yk}
J.~Ben~Geloun, \emph{{Two and four-loop {$\beta$}-functions of rank 4
  renormalizable tensor field theories}},
  \href{https://doi.org/10.1088/0264-9381/29/23/235011}{\emph{Class. Quant.
  Grav.} {\bfseries 29} (2012) 235011}
  [\href{https://arxiv.org/abs/1205.5513}{{\ttfamily 1205.5513}}].

\bibitem{Carrozza:2012uv}
S.~Carrozza, D.~Oriti and V.~Rivasseau, \emph{{Renormalization of Tensorial
  Group Field Theories: {Abelian} {$U(1)$} Models in Four Dimensions}},
  \href{https://doi.org/10.1007/s00220-014-1954-8}{\emph{Commun. Math. Phys.}
  {\bfseries 327} (2014) 603}
  [\href{https://arxiv.org/abs/1207.6734}{{\ttfamily 1207.6734}}].

\bibitem{Geloun:2014kpa}
J.~Ben~Geloun, \emph{{Renormalizable Models in Rank {$d\geq 2$} Tensorial Group
  Field Theory}},
  \href{https://doi.org/10.1007/s00220-014-2142-6}{\emph{Commun. Math. Phys.}
  {\bfseries 332} (2014) 117}
  [\href{https://arxiv.org/abs/1306.1201}{{\ttfamily 1306.1201}}].

\bibitem{Lahoche:2015ola}
V.~Lahoche, D.~Oriti and V.~Rivasseau, \emph{{Renormalization of an {Abelian}
  Tensor Group Field Theory: Solution at Leading Order}},
  \href{https://arxiv.org/abs/1501.02086}{{\ttfamily 1501.02086}}.

\bibitem{Benedetti:2014qsa}
D.~Benedetti, J.~Ben~Geloun and D.~Oriti, \emph{{Functional Renormalisation
  Group Approach for Tensorial Group Field Theory: a Rank-3 Model}},
  \href{https://doi.org/10.1007/JHEP03(2015)084}{\emph{JHEP} {\bfseries 03}
  (2015) 084} [\href{https://arxiv.org/abs/1411.3180}{{\ttfamily 1411.3180}}].

\bibitem{Lahoche:2018hou}
V.~Lahoche and D.~O. Samary, \emph{{Progress in the solving nonperturbative
  renormalization group for tensorial group field theory}},
  \href{https://doi.org/10.3390/universe5030086}{\emph{Universe} {\bfseries 5}
  (2019) 86} [\href{https://arxiv.org/abs/1812.00905}{{\ttfamily 1812.00905}}].

\bibitem{Brezin:1992yc}
E.~Brezin and J.~Zinn-Justin, \emph{{Renormalization group approach to matrix
  models}}, \href{https://doi.org/10.1016/0370-2693(92)91953-7}{\emph{Phys.\
  Lett.\ B} {\bfseries 288} (1992) 54}
  [\href{https://arxiv.org/abs/hep-th/9206035}{{\ttfamily hep-th/9206035}}].

\bibitem{Ambjorn:1990ge}
J.~Ambjorn, B.~Durhuus and T.~Jonsson, \emph{{Three-dimensional simplicial
  quantum gravity and generalized matrix models}},
  \href{https://doi.org/10.1142/S0217732391001184}{\emph{Mod. Phys. Lett.}
  {\bfseries A6} (1991) 1133}.

\bibitem{Sasakura:1990fs}
N.~Sasakura, \emph{{Tensor model for gravity and orientability of manifold}},
  \href{https://doi.org/10.1142/S0217732391003055}{\emph{Mod. Phys. Lett.}
  {\bfseries A6} (1991) 2613}.

\bibitem{Gross:1991hx}
M.~Gross, \emph{{Tensor models and simplicial quantum gravity in > 2-D}},
  \href{https://doi.org/10.1016/S0920-5632(05)80015-5}{\emph{Nucl.\ Phys.\ B
  Proc.\ Suppl.} {\bfseries 25A} (1992) 144}.

\bibitem{Gurau:2010ba}
R.~Gurau, \emph{{The 1/N expansion of colored tensor models}},
  \href{https://doi.org/10.1007/s00023-011-0101-8}{\emph{Annales Henri
  Poincare} {\bfseries 12} (2011) 829}
  [\href{https://arxiv.org/abs/1011.2726}{{\ttfamily 1011.2726}}].

\bibitem{Gurau:2011aq}
R.~Gurau and V.~Rivasseau, \emph{{The 1/N expansion of colored tensor models in
  arbitrary dimension}},
  \href{https://doi.org/10.1209/0295-5075/95/50004}{\emph{Europhys.\ Lett.}
  {\bfseries 95} (2011) 50004}
  [\href{https://arxiv.org/abs/1101.4182}{{\ttfamily 1101.4182}}].

\bibitem{Gurau:2011xq}
R.~Gurau, \emph{{The complete 1/N expansion of colored tensor models in
  arbitrary dimension}},
  \href{https://doi.org/10.1007/s00023-011-0118-z}{\emph{Annales Henri
  Poincare} {\bfseries 13} (2012) 399}
  [\href{https://arxiv.org/abs/1102.5759}{{\ttfamily 1102.5759}}].

\bibitem{Bonzom:2011zz}
V.~Bonzom, R.~Gurau, A.~Riello and V.~Rivasseau, \emph{{Critical behavior of
  colored tensor models in the large N limit}},
  \href{https://doi.org/10.1016/j.nuclphysb.2011.07.022}{\emph{Nucl.\ Phys.\ B}
  {\bfseries 853} (2011) 174}
  [\href{https://arxiv.org/abs/1105.3122}{{\ttfamily 1105.3122}}].

\bibitem{Bonzom:2012hw}
V.~Bonzom, R.~Gurau and V.~Rivasseau, \emph{{Random tensor models in the large
  N limit: Uncoloring the colored tensor models}},
  \href{https://doi.org/10.1103/PhysRevD.85.084037}{\emph{Phys.\ Rev.\ D}
  {\bfseries 85} (2012) 084037}
  [\href{https://arxiv.org/abs/1202.3637}{{\ttfamily 1202.3637}}].

\bibitem{Carrozza:2015adg}
S.~Carrozza and A.~Tanasa, \emph{{$O(N)$} random tensor models},
  \href{https://doi.org/10.1007/s11005-016-0879-x}{\emph{Lett. Math. Phys.}
  {\bfseries 106} (2016) 1531}
  [\href{https://arxiv.org/abs/1512.06718}{{\ttfamily 1512.06718}}].

\bibitem{Prakash:2019zia}
S.~Prakash and R.~Sinha, \emph{Melonic dominance in subchromatic sextic tensor
  models},  \href{https://arxiv.org/abs/1908.07178}{{\ttfamily 1908.07178}}.

\bibitem{Tanasa:2015uhr}
A.~Tanasa, \emph{The multi-orientable random tensor model, a review},
  \href{https://doi.org/10.3842/SIGMA.2016.056}{\emph{SIGMA} {\bfseries 12}
  (2016) 056} [\href{https://arxiv.org/abs/1512.02087}{{\ttfamily
  1512.02087}}].

\bibitem{Klebanov:2017nlk}
I.~R. Klebanov and G.~Tarnopolsky, \emph{On large {$N$} limit of symmetric
  traceless tensor models},
  \href{https://doi.org/10.1007/JHEP10(2017)037}{\emph{JHEP} {\bfseries 10}
  (2017) 037} [\href{https://arxiv.org/abs/1706.00839}{{\ttfamily
  1706.00839}}].

\bibitem{Benedetti:2017qxl}
D.~Benedetti, S.~Carrozza, R.~Gurau and M.~Kolanowski, \emph{{The $1/N$
  expansion of the symmetric traceless and the antisymmetric tensor models in
  rank three}}, \href{https://doi.org/10.1007/s00220-019-03551-z}{\emph{Commun.
  Math. Phys.} {\bfseries 371} (2019) 55}
  [\href{https://arxiv.org/abs/1712.00249}{{\ttfamily 1712.00249}}].

\bibitem{Carrozza:2018ewt}
S.~Carrozza, \emph{{Large $N$ limit of irreducible tensor models: $O(N)$
  rank-$3$ tensors with mixed permutation symmetry}},
  \href{https://doi.org/10.1007/JHEP06(2018)039}{\emph{JHEP} {\bfseries 06}
  (2018) 039} [\href{https://arxiv.org/abs/1803.02496}{{\ttfamily
  1803.02496}}].

\bibitem{Sachdev:1992fk}
S.~Sachdev and J.~Ye, \emph{{Gapless spin fluid ground state in a random,
  quantum Heisenberg magnet}},
  \href{https://doi.org/10.1103/PhysRevLett.70.3339}{\emph{Phys. Rev. Lett.}
  {\bfseries 70} (1993) 3339}
  [\href{https://arxiv.org/abs/cond-mat/9212030}{{\ttfamily
  cond-mat/9212030}}].

\bibitem{Kitaev}
A.~Kitaev, \emph{{A simple model of quantum holography}}, {\emph{KITP strings
  seminar and Entanglement 2015} (Feb. 12, April 7, and May 27, 2015) }.

\bibitem{Polchinski:2016xgd}
J.~Polchinski and V.~Rosenhaus, \emph{{The Spectrum in the Sachdev-Ye-Kitaev
  Model}}, \href{https://doi.org/10.1007/JHEP04(2016)001}{\emph{JHEP}
  {\bfseries 04} (2016) 001}
  [\href{https://arxiv.org/abs/1601.06768}{{\ttfamily 1601.06768}}].

\bibitem{Maldacena:2016hyu}
J.~Maldacena and D.~Stanford, \emph{{Remarks on the Sachdev-Ye-Kitaev model}},
  \href{https://doi.org/10.1103/PhysRevD.94.106002}{\emph{Phys. Rev.}
  {\bfseries D94} (2016) 106002}
  [\href{https://arxiv.org/abs/1604.07818}{{\ttfamily 1604.07818}}].

\bibitem{Witten:2016iux}
E.~Witten, \emph{An {SYK}-like model without disorder},
  \href{https://doi.org/10.1088/1751-8121/ab3752}{\emph{J. Phys.} {\bfseries
  A52} (2019) 474002} [\href{https://arxiv.org/abs/1610.09758}{{\ttfamily
  1610.09758}}].

\bibitem{Klebanov:2016xxf}
I.~R. Klebanov and G.~Tarnopolsky, \emph{Uncolored random tensors, melon
  diagrams, and the {SYK} models},
  \href{https://doi.org/10.1103/PhysRevD.95.046004}{\emph{Phys. Rev.}
  {\bfseries D95} (2017) 046004}
  [\href{https://arxiv.org/abs/1611.08915}{{\ttfamily 1611.08915}}].

\bibitem{Peng:2016mxj}
C.~Peng, M.~Spradlin and A.~Volovich, \emph{{A Supersymmetric SYK-like Tensor
  Model}}, \href{https://doi.org/10.1007/JHEP05(2017)062}{\emph{JHEP}
  {\bfseries 05} (2017) 062}
  [\href{https://arxiv.org/abs/1612.03851}{{\ttfamily 1612.03851}}].

\bibitem{Krishnan:2016bvg}
C.~Krishnan, S.~Sanyal and P.~N. Bala~Subramanian, \emph{Quantum chaos and
  holographic tensor models},
  \href{https://doi.org/10.1007/JHEP03(2017)056}{\emph{JHEP} {\bfseries 03}
  (2017) 056} [\href{https://arxiv.org/abs/1612.06330}{{\ttfamily
  1612.06330}}].

\bibitem{Bonzom:2017pqs}
V.~Bonzom, L.~Lionni and A.~Tanasa, \emph{{Diagrammatics of a colored SYK model
  and of an SYK-like tensor model, leading and next-to-leading orders}},
  \href{https://doi.org/10.1063/1.4983562}{\emph{J. Math. Phys.} {\bfseries 58}
  (2017) 052301} [\href{https://arxiv.org/abs/1702.06944}{{\ttfamily
  1702.06944}}].

\bibitem{Beccaria:2017aqc}
M.~Beccaria and A.~A. Tseytlin, \emph{{Partition function of free conformal
  fields in 3-plet representation}},
  \href{https://doi.org/10.1007/JHEP05(2017)053}{\emph{JHEP} {\bfseries 05}
  (2017) 053} [\href{https://arxiv.org/abs/1703.04460}{{\ttfamily
  1703.04460}}].

\bibitem{Gubser:2017qed}
S.~S. Gubser, M.~Heydeman, C.~Jepsen, S.~Parikh, I.~Saberi, B.~Stoica et~al.,
  \emph{{Melonic theories over diverse number systems}},
  \href{https://doi.org/10.1103/PhysRevD.98.126007}{\emph{Phys.\ Rev.\ D}
  {\bfseries 98} (2018) 126007}
  [\href{https://arxiv.org/abs/1707.01087}{{\ttfamily 1707.01087}}].

\bibitem{Giombi:2017dtl}
S.~Giombi, I.~R. Klebanov and G.~Tarnopolsky, \emph{{Bosonic tensor models at
  large {$N$} and small {$\epsilon$}}},
  \href{https://doi.org/10.1103/PhysRevD.96.106014}{\emph{Phys. Rev.}
  {\bfseries D96} (2017) 106014}
  [\href{https://arxiv.org/abs/1707.03866}{{\ttfamily 1707.03866}}].

\bibitem{Bulycheva:2017ilt}
K.~Bulycheva, I.~R. Klebanov, A.~Milekhin and G.~Tarnopolsky, \emph{Spectra of
  operators in large {$N$} tensor models},
  \href{https://doi.org/10.1103/PhysRevD.97.026016}{\emph{Phys.\ Rev.\ D}
  {\bfseries 97} (2018) 026016}
  [\href{https://arxiv.org/abs/1707.09347}{{\ttfamily 1707.09347}}].

\bibitem{Choudhury:2017tax}
S.~Choudhury, A.~Dey, I.~Halder, L.~Janagal, S.~Minwalla and R.~Poojary,
  \emph{{Notes on melonic $O(N)^{q-1}$ tensor models}},
  \href{https://doi.org/10.1007/JHEP06(2018)094}{\emph{JHEP} {\bfseries 06}
  (2018) 094} [\href{https://arxiv.org/abs/1707.09352}{{\ttfamily
  1707.09352}}].

\bibitem{Yoon:2017nig}
J.~Yoon, \emph{{SYK} models and {SYK}-like tensor models with global symmetry},
  \href{https://doi.org/10.1007/JHEP10(2017)183}{\emph{JHEP} {\bfseries 10}
  (2017) 183} [\href{https://arxiv.org/abs/1707.01740}{{\ttfamily
  1707.01740}}].

\bibitem{Krishnan:2017lra}
C.~Krishnan, K.~V. Pavan~Kumar and D.~Rosa, \emph{{Contrasting SYK-like
  Models}}, \href{https://doi.org/10.1007/JHEP01(2018)064}{\emph{JHEP}
  {\bfseries 01} (2018) 064}
  [\href{https://arxiv.org/abs/1709.06498}{{\ttfamily 1709.06498}}].

\bibitem{Prakash:2017hwq}
S.~Prakash and R.~Sinha, \emph{A complex fermionic tensor model in $d$
  dimensions}, \href{https://doi.org/10.1007/JHEP02(2018)086}{\emph{JHEP}
  {\bfseries 02} (2018) 086}
  [\href{https://arxiv.org/abs/1710.09357}{{\ttfamily 1710.09357}}].

\bibitem{Benedetti:2017fmp}
D.~Benedetti, S.~Carrozza, R.~Gurau and A.~Sfondrini, \emph{{Tensorial
  Gross-Neveu models}},
  \href{https://doi.org/10.1007/JHEP01(2018)003}{\emph{JHEP} {\bfseries 01}
  (2018) 003} [\href{https://arxiv.org/abs/1710.10253}{{\ttfamily
  1710.10253}}].

\bibitem{Chang:2018sve}
C.-M. Chang, S.~Colin-Ellerin and M.~Rangamani, \emph{{On Melonic Supertensor
  Models}}, \href{https://doi.org/10.1007/JHEP10(2018)157}{\emph{JHEP}
  {\bfseries 10} (2018) 157}
  [\href{https://arxiv.org/abs/1806.09903}{{\ttfamily 1806.09903}}].

\bibitem{Giombi:2018qgp}
S.~Giombi, I.~R. Klebanov, F.~Popov, S.~Prakash and G.~Tarnopolsky,
  \emph{Prismatic large {$N$} models for bosonic tensors},
  \href{https://doi.org/10.1103/PhysRevD.98.105005}{\emph{Phys. Rev.}
  {\bfseries D98} (2018) 105005}
  [\href{https://arxiv.org/abs/1808.04344}{{\ttfamily 1808.04344}}].

\bibitem{Benedetti:2018ghn}
D.~Benedetti and N.~Delporte, \emph{{Phase diagram and fixed points of
  tensorial Gross-Neveu models in three dimensions}},
  \href{https://doi.org/10.1007/JHEP01(2019)218}{\emph{JHEP} {\bfseries 01}
  (2019) 218} [\href{https://arxiv.org/abs/1810.04583}{{\ttfamily
  1810.04583}}].

\bibitem{Kim:2019upg}
J.~Kim, I.~R. Klebanov, G.~Tarnopolsky and W.~Zhao, \emph{Symmetry breaking in
  coupled {SYK} or tensor models},
  \href{https://doi.org/10.1103/PhysRevX.9.021043}{\emph{Phys.\ Rev.\ X}
  {\bfseries 9} (2019) 021043}
  [\href{https://arxiv.org/abs/1902.02287}{{\ttfamily 1902.02287}}].

\bibitem{Benedetti:2019eyl}
D.~Benedetti, R.~Gurau and S.~Harribey, \emph{{Line of fixed points in a
  bosonic tensor model}},
  \href{https://doi.org/10.1007/JHEP06(2019)053}{\emph{JHEP} {\bfseries 06}
  (2019) 053} [\href{https://arxiv.org/abs/1903.03578}{{\ttfamily
  1903.03578}}].

\bibitem{Klebanov:2019jup}
I.~R. Klebanov, P.~N. Pallegar and F.~K. Popov, \emph{{Majorana Fermion Quantum
  Mechanics for Higher Rank Tensors}},
  \href{https://doi.org/10.1103/PhysRevD.100.086003}{\emph{Phys. Rev. D}
  {\bfseries 100} (2019) 086003}
  [\href{https://arxiv.org/abs/1905.06264}{{\ttfamily 1905.06264}}].

\bibitem{Popov:2019nja}
F.~K. Popov, \emph{{Supersymmetric tensor model at large $N$ and small
  $\epsilon$}}, \href{https://doi.org/10.1103/PhysRevD.101.026020}{\emph{Phys.\
  Rev.\ D} {\bfseries 101} (2020) 026020}
  [\href{https://arxiv.org/abs/1907.02440}{{\ttfamily 1907.02440}}].

\bibitem{Benedetti:2019ikb}
D.~Benedetti, R.~Gurau, S.~Harribey and K.~Suzuki, \emph{{Hints of unitarity at
  large $N$ in the $O(N)^3$ tensor field theory}},
  \href{https://doi.org/10.1007/JHEP02(2020)072}{\emph{JHEP} {\bfseries 02}
  (2020) 072} [\href{https://arxiv.org/abs/1909.07767}{{\ttfamily
  1909.07767}}].

\bibitem{Benedetti:2019rja}
D.~Benedetti, N.~Delporte, S.~Harribey and R.~Sinha, \emph{{Sextic tensor field
  theories in rank $3$ and $5$}},
  \href{https://arxiv.org/abs/1912.06641}{{\ttfamily 1912.06641}}.

\bibitem{Benedetti:2020yvb}
D.~Benedetti, R.~Gurau and K.~Suzuki, \emph{Conformal symmetry and composite
  operators in the {$O(N)^3$} tensor field theory},
  \href{https://arxiv.org/abs/2002.07652}{{\ttfamily 2002.07652}}.

\bibitem{Klebanov:2018fzb}
I.~R. Klebanov, F.~Popov and G.~Tarnopolsky, \emph{{TASI} lectures on large
  {$N$} tensor models}, \href{https://doi.org/10.22323/1.305.0004}{\emph{PoS}
  {\bfseries TASI2017} (2018) 004}
  [\href{https://arxiv.org/abs/1808.09434}{{\ttfamily 1808.09434}}].

\bibitem{Patashinskii:1964}
A.~Z. Patashinskii and V.~L. Pokrovskii, \emph{Second order phase transitions
  in a bose fluid},
  \href{https://doi.org/http://www.jetp.ac.ru/cgi-bin/e/index/e/19/3/p677?a=list}{\emph{JETP}
  {\bfseries 19} (1964) 677}.

\bibitem{Gribov:1968fg}
V.~Gribov and A.~A. Migdal, \emph{Strong coupling in the {Pomeranchuk} pole
  problem},
  \href{https://doi.org/http://www.jetp.ac.ru/cgi-bin/e/index/e/28/4/p784?a=list}{\emph{JETP}
  {\bfseries 28} (1969) 784}.

\bibitem{Blaizot:2003br}
J.-P. Blaizot, E.~Iancu and U.~Reinosa, \emph{{Renormalizability of Phi
  derivable approximations in scalar phi**4 theory}},
  \href{https://doi.org/10.1016/j.physletb.2003.06.008}{\emph{Phys. Lett.}
  {\bfseries B568} (2003) 160}
  [\href{https://arxiv.org/abs/hep-ph/0301201}{{\ttfamily hep-ph/0301201}}].

\bibitem{Berges:2005hc}
J.~Berges, S.~Borsanyi, U.~Reinosa and J.~Serreau, \emph{{Nonperturbative
  renormalization for 2PI effective action techniques}},
  \href{https://doi.org/10.1016/j.aop.2005.06.001}{\emph{Annals Phys.}
  {\bfseries 320} (2005) 344}
  [\href{https://arxiv.org/abs/hep-ph/0503240}{{\ttfamily hep-ph/0503240}}].

\bibitem{Blaizot:2010zx}
J.-P. Blaizot, J.~M. Pawlowski and U.~Reinosa, \emph{{Exact renormalization
  group and $\Phi$-derivable approximations}},
  \href{https://doi.org/10.1016/j.physletb.2010.12.058}{\emph{Phys. Lett.}
  {\bfseries B696} (2011) 523}
  [\href{https://arxiv.org/abs/1009.6048}{{\ttfamily 1009.6048}}].

\bibitem{Benedetti:2018goh}
D.~Benedetti and R.~Gurau, \emph{{2PI effective action for the SYK model and
  tensor field theories}},
  \href{https://doi.org/10.1007/JHEP05(2018)156}{\emph{JHEP} {\bfseries 05}
  (2018) 156} [\href{https://arxiv.org/abs/1802.05500}{{\ttfamily
  1802.05500}}].

\bibitem{Fisher:1972zz}
M.~E. Fisher, S.-k. Ma and B.~Nickel, \emph{Critical exponents for long-range
  interactions},
  \href{https://doi.org/10.1103/PhysRevLett.29.917}{\emph{Phys.Rev.Lett.}
  {\bfseries 29} (1972) 917}.

\bibitem{Sak:1973}
J.~Sak, \emph{Recursion relations and fixed points for ferromagnets with
  long-range interactions},
  \href{https://doi.org/10.1103/PhysRevB.8.281}{\emph{Phys. Rev. B} {\bfseries
  8} (1973) 281}.

\bibitem{Brydges:2002wq}
D.~C. Brydges, P.~K. Mitter and B.~Scoppola, \emph{{Critical (Phi**4)(3,
  epsilon)}}, \href{https://doi.org/10.1007/s00220-003-0895-4}{\emph{Commun.
  Math. Phys.} {\bfseries 240} (2003) 281}
  [\href{https://arxiv.org/abs/hep-th/0206040}{{\ttfamily hep-th/0206040}}].

\bibitem{Abdesselam:2006qg}
A.~Abdesselam, \emph{A complete renormalization group trajectory between two
  fixed points},
  \href{https://doi.org/10.1007/s00220-007-0352-x}{\emph{Commun.Math.Phys.}
  {\bfseries 276} (2007) 727}
  [\href{https://arxiv.org/abs/math-ph/0610018}{{\ttfamily math-ph/0610018}}].

\bibitem{Brezin:2014}
E.~Brezin, G.~Parisi and F.~Ricci-Tersenghi, \emph{The crossover region between
  long-range and short-range interactions for the critical exponents},
  \href{https://doi.org/10.1007/s10955-014-1081-0}{\emph{J. Stat. Phys.}
  {\bfseries 157} (2014) 855}
  [\href{https://arxiv.org/abs/1407.3358}{{\ttfamily 1407.3358}}].

\bibitem{Defenu:2014}
N.~Defenu, A.~Trombettoni and A.~Codello, \emph{{Fixed-point structure and
  effective fractional dimensionality for $O(N)$ models with long-range
  interactions}}, \href{https://doi.org/10.1103/physreve.92.052113}{\emph{Phys.
  Rev.} {\bfseries E92} (2015) 052113}
  [\href{https://arxiv.org/abs/1409.8322}{{\ttfamily 1409.8322}}].

\bibitem{Paulos:2015jfa}
M.~F. Paulos, S.~Rychkov, B.~C. van Rees and B.~Zan, \emph{Conformal invariance
  in the long-range ising model},
  \href{https://doi.org/10.1016/j.nuclphysb.2015.10.018}{\emph{Nucl.Phys.B}
  {\bfseries 902} (2016) 246}
  [\href{https://arxiv.org/abs/1509.00008}{{\ttfamily 1509.00008}}].

\bibitem{Behan:2017emf}
C.~Behan, L.~Rastelli, S.~Rychkov and B.~Zan, \emph{A scaling theory for the
  long-range to short-range crossover and an infrared duality},
  \href{https://doi.org/10.1088/1751-8121/aa8099}{\emph{J.Phys.A} {\bfseries
  50} (2017) 354002} [\href{https://arxiv.org/abs/1703.05325}{{\ttfamily
  1703.05325}}].

\bibitem{RTM}
R.~Gurau, \emph{{Random Tensors}}. Oxford University Press, Oxford, 2016.

\bibitem{Lionni:2017yvi}
L.~Lionni, \emph{{Colored discrete spaces: higher dimensional combinatorial
  maps and quantum gravity}}, Ph.D. thesis, Saclay, 2017.
\newblock \href{https://arxiv.org/abs/1710.03663}{{\ttfamily 1710.03663}}.
\newblock 10.1007/978-3-319-96023-4.

\bibitem{Ferrari:2017jgw}
F.~Ferrari, V.~Rivasseau and G.~Valette, \emph{A new large {N} expansion for
  general matrix-tensor models},
  \href{https://arxiv.org/abs/1709.07366}{{\ttfamily 1709.07366}}.

\bibitem{Gubser:2018yec}
S.~S. Gubser, C.~Jepsen, Z.~Ji and B.~Trundy, \emph{{Higher melonic theories}},
  \href{https://doi.org/10.1007/JHEP09(2018)049}{\emph{JHEP} {\bfseries 09}
  (2018) 049} [\href{https://arxiv.org/abs/1806.04800}{{\ttfamily
  1806.04800}}].

\bibitem{Geloun:2013kta}
J.~Ben~Geloun and S.~Ramgoolam, \emph{Counting tensor model observables and
  branched covers of the 2-sphere},
  \href{https://arxiv.org/abs/1307.6490}{{\ttfamily 1307.6490}}.

\bibitem{BenGeloun:2017vwn}
J.~Ben~Geloun and S.~Ramgoolam, \emph{Tensor models, {Kronecker} coefficients
  and permutation centralizer algebras},
  \href{https://doi.org/10.1007/JHEP11(2017)092}{\emph{JHEP} {\bfseries 11}
  (2017) 092} [\href{https://arxiv.org/abs/1708.03524}{{\ttfamily
  1708.03524}}].

\bibitem{Avohou:2019qrl}
R.~C. Avohou, J.~Ben~Geloun and N.~Dub, \emph{{On the counting of $O(N)$ tensor
  invariants}},  \href{https://arxiv.org/abs/1907.04668}{{\ttfamily
  1907.04668}}.

\bibitem{Cornwall:1974vz}
J.~M. Cornwall, R.~Jackiw and E.~Tomboulis, \emph{Effective action for
  composite operators},
  \href{https://doi.org/10.1103/PhysRevD.10.2428}{\emph{Phys. Rev.} {\bfseries
  D10} (1974) 2428}.

\bibitem{Berges:2004yj}
J.~Berges, \emph{{Introduction to nonequilibrium quantum field theory}},
  \href{https://doi.org/10.1063/1.1843591}{\emph{AIP Conf. Proc.} {\bfseries
  739} (2005) 3} [\href{https://arxiv.org/abs/hep-ph/0409233}{{\ttfamily
  hep-ph/0409233}}].

\bibitem{Gurau:2019qag}
R.~Gurau, \emph{Notes on tensor models and tensor field theories},
  \href{https://arxiv.org/abs/1907.03531}{{\ttfamily 1907.03531}}.

\bibitem{Benedetti:2019sop}
D.~Benedetti and I.~Costa, \emph{{$SO(3)$-invariant phase of the $O(N)^3$
  tensor model}},  \href{https://arxiv.org/abs/1912.07311}{{\ttfamily
  1912.07311}}.

\bibitem{Aarts:2002dj}
G.~Aarts, D.~Ahrensmeier, R.~Baier, J.~Berges and J.~Serreau, \emph{{Far from
  equilibrium dynamics with broken symmetries from the 2PI - 1/N expansion}},
  \href{https://doi.org/10.1103/PhysRevD.66.045008}{\emph{Phys. Rev.}
  {\bfseries D66} (2002) 045008}
  [\href{https://arxiv.org/abs/hep-ph/0201308}{{\ttfamily hep-ph/0201308}}].

\bibitem{Berges:2001fi}
J.~Berges, \emph{{Controlled nonperturbative dynamics of quantum fields
  out-of-equilibrium}},
  \href{https://doi.org/10.1016/S0375-9474(01)01295-7}{\emph{Nucl. Phys.}
  {\bfseries A699} (2002) 847}
  [\href{https://arxiv.org/abs/hep-ph/0105311}{{\ttfamily hep-ph/0105311}}].

\bibitem{Liu:2018jhs}
J.~Liu, E.~Perlmutter, V.~Rosenhaus and D.~Simmons-Duffin,
  \emph{{$d$-dimensional SYK, AdS Loops, and $6j$ Symbols}},
  \href{https://doi.org/10.1007/JHEP03(2019)052}{\emph{JHEP} {\bfseries 03}
  (2019) 052} [\href{https://arxiv.org/abs/1808.00612}{{\ttfamily
  1808.00612}}].

\bibitem{Simmons-Duffin:2017nub}
D.~Simmons-Duffin, D.~Stanford and E.~Witten, \emph{{A spacetime derivation of
  the Lorentzian OPE inversion formula}},
  \href{https://doi.org/10.1007/JHEP07(2018)085}{\emph{JHEP} {\bfseries 07}
  (2018) 085} [\href{https://arxiv.org/abs/1711.03816}{{\ttfamily
  1711.03816}}].

\bibitem{Wilson:1971dc}
K.~G. Wilson and M.~E. Fisher, \emph{{Critical exponents in 3.99 dimensions}},
  \href{https://doi.org/10.1103/PhysRevLett.28.240}{\emph{Phys.\ Rev.\ Lett.}
  {\bfseries 28} (1972) 240}.

\bibitem{Jack:1990eb}
I.~Jack and H.~Osborn, \emph{Analogs for the $c$ theorem for four-dimensional
  renormalizable field theories},
  \href{https://doi.org/10.1016/0550-3213(90)90584-Z}{\emph{Nucl.\ Phys.\ B}
  {\bfseries 343} (1990) 647}.

\bibitem{Breitenlohner:1982jf}
P.~Breitenlohner and D.~Z. Freedman, \emph{Stability in gauged extended
  supergravity},
  \href{https://doi.org/10.1016/0003-4916(82)90116-6}{\emph{Annals Phys.}
  {\bfseries 144} (1982) 249}.

\bibitem{Gross:2017vhb}
D.~J. Gross and V.~Rosenhaus, \emph{{A line of CFTs: from generalized free
  fields to SYK}}, \href{https://doi.org/10.1007/JHEP07(2017)086}{\emph{JHEP}
  {\bfseries 07} (2017) 086}
  [\href{https://arxiv.org/abs/1706.07015}{{\ttfamily 1706.07015}}].

\bibitem{Fisher:1978pf}
M.~E. Fisher, \emph{{Yang-Lee} edge singularity and phi**3 field theory},
  \href{https://doi.org/10.1103/PhysRevLett.40.1610}{\emph{Phys. Rev. Lett.}
  {\bfseries 40} (1978) 1610}.

\bibitem{Cardy:1985yy}
J.~L. Cardy, \emph{Conformal invariance and the {Yang-Lee} edge singularity in
  two-dimensions},
  \href{https://doi.org/10.1103/PhysRevLett.54.1354}{\emph{Phys. Rev. Lett.}
  {\bfseries 54} (1985) 1354}.

\bibitem{Poland:2018epd}
D.~Poland, S.~Rychkov and A.~Vichi, \emph{The conformal bootstrap: Theory,
  numerical techniques, and applications},
  \href{https://doi.org/10.1103/RevModPhys.91.015002}{\emph{Rev.Mod.Phys.}
  {\bfseries 91} (2019) 015002}
  [\href{https://arxiv.org/abs/1805.04405}{{\ttfamily 1805.04405}}].

\bibitem{Delporte:2018iyf}
N.~Delporte and V.~Rivasseau, \emph{The tensor track {V}: Holographic tensors},
   in \emph{{Proceedings, 17th Hellenic School and Workshops on Elementary
  Particle Physics and Gravity (CORFU2017)}: {Corfu, Greece, September 2-28,
  2017}}, 4, 2018, \href{https://arxiv.org/abs/1804.11101}{{\ttfamily
  1804.11101}}.

\bibitem{Bondi:1989nq}
A.~Bondi, G.~Curci, G.~Paffuti and P.~Rossi, \emph{Metric and central charge in
  the perturbative approach to two-dimensional fermionic models},
  \href{https://doi.org/10.1016/0003-4916(90)90380-7}{\emph{Annals Phys.}
  {\bfseries 199} (1990) 268}.

\bibitem{Witten:1978qu}
E.~Witten, \emph{{Chiral symmetry, the 1/N expansion, and the SU(N) Thirring
  model}}, \href{https://doi.org/10.1016/0550-3213(78)90416-9}{\emph{Nucl.\
  Phys.\ B} {\bfseries 145} (1978) 110}.

\bibitem{Gawedzki:1985ed}
K.~Gawedzki and A.~Kupiainen, \emph{Renormalizing the nonrenormalizable},
  \href{https://doi.org/10.1103/PhysRevLett.55.363}{\emph{Phys. Rev. Lett.}
  {\bfseries 55} (1985) 363}.

\bibitem{Gawedzki:1985jn}
K.~Gawedzki and A.~Kupiainen, \emph{Renormalization of a nonrenormalizable
  quantum field theory},
  \href{https://doi.org/10.1016/0550-3213(85)90062-8}{\emph{Nucl. Phys. B}
  {\bfseries 262} (1985) 33}.

\bibitem{Kazakov:2018qbr}
V.~Kazakov and E.~Olivucci, \emph{Biscalar integrable conformal field theories
  in any dimension},
  \href{https://doi.org/10.1103/PhysRevLett.121.131601}{\emph{Phys.Rev.Lett.}
  {\bfseries 121} (2018) 131601}
  [\href{https://arxiv.org/abs/1801.09844}{{\ttfamily 1801.09844}}].

\bibitem{Gurdogan:2015csr}
O.~Gurdogan and V.~Kazakov, \emph{New integrable 4d quantum field theories from
  strongly deformed planar {$\mathcal N = 4$} supersymmetric {Yang-Mills}
  theory},
  \href{https://doi.org/10.1103/PhysRevLett.117.201602}{\emph{Phys.Rev.Lett.}
  {\bfseries 117} (2016) 201602}
  [\href{https://arxiv.org/abs/1512.06704}{{\ttfamily 1512.06704}}].

\bibitem{Ferrari:2017ryl}
F.~Ferrari, \emph{The large {D} limit of planar diagrams},
  \href{https://doi.org/10.4171/AIHPD/76}{\emph{Ann. Inst. Henri Poincar\'e
  Comb. Phys. Interact.} {\bfseries 6} (2019) 427}
  [\href{https://arxiv.org/abs/1701.01171}{{\ttfamily 1701.01171}}].

\bibitem{Azeyanagi:2017drg}
T.~Azeyanagi, F.~Ferrari and F.~I. Schaposnik~Massolo, \emph{Phase diagram of
  planar matrix quantum mechanics, tensor, and {Sachdev-Ye-Kitaev} models},
  \href{https://doi.org/10.1103/PhysRevLett.120.061602}{\emph{Phys. Rev. Lett.}
  {\bfseries 120} (2018) 061602}
  [\href{https://arxiv.org/abs/1707.03431}{{\ttfamily 1707.03431}}].

\bibitem{Azeyanagi:2017mre}
T.~Azeyanagi, F.~Ferrari, P.~Gregori, L.~Leduc and G.~Valette, \emph{More on
  the new large {$D$} limit of matrix models},
  \href{https://doi.org/10.1016/j.aop.2018.04.010}{\emph{Annals Phys.}
  {\bfseries 393} (2018) 308}
  [\href{https://arxiv.org/abs/1710.07263}{{\ttfamily 1710.07263}}].

\bibitem{Benedetti:2020iyz}
D.~Benedetti, S.~Carrozza, R.~Toriumi and G.~Valette, \emph{{Multiple scaling
  limits of $\mathrm{U}(N)^2 \times \mathrm{O}(D)$ multi-matrix models}},
  \href{https://arxiv.org/abs/2003.02100}{{\ttfamily 2003.02100}}.

\bibitem{Carrozza:2020eaz}
S.~Carrozza, F.~Ferrari, A.~Tanasa and G.~Valette, \emph{{On the large $D$
  expansion of Hermitian multi-matrix models}},
  \href{https://arxiv.org/abs/2003.04152}{{\ttfamily 2003.04152}}.

\bibitem{Aharony:2008ug}
O.~Aharony, O.~Bergman, D.~L. Jafferis and J.~Maldacena, \emph{{N=6
  superconformal Chern-Simons-matter theories, M2-branes and their gravity
  duals}}, \href{https://doi.org/10.1088/1126-6708/2008/10/091}{\emph{JHEP}
  {\bfseries 10} (2008) 091} [\href{https://arxiv.org/abs/0806.1218}{{\ttfamily
  0806.1218}}].

\end{thebibliography}
\end{document}